\documentclass[final]{elsart}

\usepackage{graphicx}
\usepackage{amssymb}
\usepackage{amsmath}
\usepackage{mathrsfs}
\usepackage{url}
\usepackage{bm}
\usepackage{subfigure}
\usepackage{pstricks}

\begin{document}

\begin{frontmatter}
  \title{Modeling spatiotemporal noise covariance \\ for MEG/EEG
    source analysis}
  
  \author[unm,lanl]{S. M. Plis\corauthref{pliz}}
  \corauth[pliz]{the corresponding author}
  \ead{pliz@lanl.gov}
  \author[lanl]{J. S. George}
  \author[lanl]{S. C. Jun}
  \author[lanl]{J. Par\'{e}-Blagoev}
  \author[lanl]{D. M. Ranken} 
  \author[lanl]{D. M. Schmidt}
  \author[lanl]{C. C. Wood}
  
  \address[lanl]{Biological and Quantum Physics Group, Los Alamos
    National Laboratory, MS-D454, Los Alamos, NM 87545, USA}
  
  \address[unm]{Department of Computer Science, University of New
    Mexico, Albuquerque, NM 87131, USA}
  
  \begin{abstract}
    We  propose  a  new  model for  approximating  spatiotemporal  noise
    covariance  for use  in MEG/EEG  source  analysis. Our  model is  an
    extension of an  existing model \cite{huizenga2002,deMunck2002} that
    uses a single Kronecker product of a pair of matrices - temporal and
    spatial  covariance; we  employ a  series of  Kronecker  products in
    order to  construct a better  approximation of the  full covariance.
    In contrast to the single-pair  model that assumes the same temporal
    structure for all spatial  components, the proposed model allows for
    distinct, independent time courses  at each spatial component.  This
    model   better  describes   spatially   and  temporally   correlated
    background activity.   At the same  time, inversion of the  model is
    fast which makes it useful in the inverse analysis. We have explored
    two  versions of  the model.   One  is based  on orthogonal  spatial
    components  of the  background. The  other, more  general  model, is
    based on independent spatial components.  Performance of the new and
    previous models is  compared in inverse solutions to  a large number
    of single dipole problems with simulated time courses and background
    from authentic MEG data.
  \end{abstract}
  
\end{frontmatter}

\section{Introduction}

The objective  of magnetoencephalography (MEG)/ electroencephalography
(EEG)  source  localization is  to  infer  active  brain regions  from
measurements outside of  the human head. Often, for  example in evoked
response  experiments, the  data from  individual stimulus  trials are
averaged,  time-locked  to the  stimulus  presentation. This  averaged
post-stimulus signal  is compared  with the statistical  properties of
background noise (averaged signal far from the stimulus time) and this
difference is  used to infer  the location and time-courses  of neural
activity that  is, at least on  average, generated in  response to the
given  stimulus.  Because such  inferences  are  based on  differences
between  signal and background,  it is  important to  characterize the
statistical properties of the background as accurately as possible.

The Central Limit Theorem lends  support to the common assumption that
the averaged background data  is Gaussian distributed, even though the
distribution of single trial background  may not be Gaussian.  The log
likelihood  function is a  common mathematical  expression quantifying
the likelihood that a given  model (e.g. of neural current) could have
produced  the   measured  data.   For   Gaussian,  zero-mean  averaged
background noise, the log likelihood function is given by:
\begin{equation}
  - \frac{1}{2} \sum_{ktk't'} \left[ b_{kt} - \int{\mathcal L}_{k}(x)j(x,t) dx\right]
  \mathbf{COV}^{-1}_{kt;k't'} \left[ b_{k't'} - \int{\mathcal L}_{k'}(x')j(x',t')
    dx'\right].
  \label{eqn:ML}
\end{equation}
Here $b_{kt}$ are the  averaged measurements (the data being analyzed)
at sensor  $k$ and time $t$;  $j(x,t)$ is the  neural (source) current
distribution over space $x$ and time $t$; ${\mathcal L}_{k}(x)$ is the
forward  or lead field  for sensor  $k$ --  the linear  operator which
connects source  currents to predicted measurements in  the absence of
noise.  $\mathbf{COV}$  is the  covariance of the  averaged background
activity, which  describes second order statistical  properties of the
MEG/EEG  data  in the  absence  of sources.   There  are  a number  of
different       inverse       algorithms       in       use       e.g.
\cite{HAMALAINEN-ETAL93,MosherJohnC:Muldma,SCHMIDT-ETAL-1999A,PHILLIPS-ETAL-1997,VanVeenBD:Locbea,Hamalainen1994}
but most  use the  likelihood formulation in  some way. In  all cases,
accurate  covariance  is required  to  solve  the source  localization
problem reliably.   The covariance is  commonly taken to  be diagonal,
even though there is ample  evidence that the background is correlated
over space and time \cite{Kuriki-etal-1994,Huizenga-1996,Lut98}.  This
may adversely affect the  results of inverse calculations, for example
by     biasing    the     locations    of     reconstructed    sources
\cite{JUN-ETAL-2002A}.

The sample  covariance of the  averaged background data is  related to
the sample covariance of the  single-trial background data by a simple
expression: $\mathbf{COV} = \frac{1}{M}\widehat{\mathbf{COV}}$.  Here,
$M$    is    the    number    of    trials    being    averaged    and
$\widehat{\mathbf{COV}}$ is the  sample covariance of the single-trial
background  data.  This  relation assumes  the trials  are independent
draws  from  the  single-trial  distribution  but does  not  assume  a
particular  form for this  distribution.  The  task of  estimating the
average background  covariance may  be accomplished by  estimating the
single-trial  background covariance and  scaling it  by the  number of
trials in the average.

Given that  MEG/EEG is measured in  $M$ trials, on $L$  sensors and in
$C$ time samples, let ${\bf E}_m$ be the $L\times C$ single trial noise
matrix at  trial $m$. In this  case, the conventional  way to estimate
the full covariance matrix of  dimension $N=LC$ for the averaged noise
is by
\begin{eqnarray}
  \mathbf{COV} &=& \frac1{M(M-1)} \sum_{m=1}^M (vec(\mathbf{E}_m) -
  vec(\bar{ \mathbf{E}}) )
  (vec(\mathbf{E}_m) - vec(\bar{\mathbf{E}}))^T, \label{eqn:smple_cov} \\
  \bar{\mathbf{E}}&=& \frac1M \sum_{m=1}^M \mathbf{E}_m,
  \label{eqn:COV}
\end{eqnarray}
where $vec(\mathbf{E})$ is all  the columns of $\mathbf{E}$ stacked in
a vector.  In order to simplify  notation in this paper  we use symbol
$M$ both for the number of  available noise samples and for the number
of stimuli over which averaging is done. Note that in the general case
these numbers may not be the same.

There are a number of reasons why this estimate of the full covariance
is difficult  to use and why  an approximation is  needed.  First, for
modern multi-sensor  detectors sufficient experimental  data is rarely
available  to  adequately  estimate  the large  number  of  parameters
present  in the  full  covariance  matrix. For  example,  for 35  time
samples and 121 channels and  considering the fact that the covariance
matrix  is  symmetric,  $8,969,730$  parameters should  be  determined
($(121*35)((121*35)+1)/2$).   This  far  exceeds  the amount  of  data
typically  available.    Second,  because  the   spatiotemporal  noise
covariance  matrix is  so  large,  a tremendous  amount  of memory  is
required  for its  storage.   Third, this  full  covariance is  almost
impossible  to  handle  in   the  likelihood  formulation,  since  the
computation  time of calculating  the inverse  still renders  the task
infeasible  in most  interesting cases,  even  if it  was possible  to
estimate the  covariance with the  given data.  A naive  algorithm for
matrix inversion takes $O(N^{3})$ time where $N$ is the dimensionality
of the  matrix.  Though there  are some improvements over  this result
for large matrices \cite{strassen1969,coppersmith1990}, the problem is
overwhelming  for  typical  computing  equipment and  for  interesting
values of $N$. To summarize: it is almost aways impossible to estimate
the full spatiotemporal covariance due to lack of data; in those cases
when  the estimation  is  possible, the  inversion is  computationally
hard; and in  any case the amount of storage required  is high. Due to
all these difficulties with  the estimation of the full spatiotemporal
covariance,  an  accurate approximation  is  needed.   In addition  to
addressing   the  above   three  problems   mentioned  above   a  good
approximation should capture as much  of the structure in the noise as
possible and reduce the errors in inverse solutions.

In  this  paper  we  describe  three different  models  of  increasing
complexity     and     different     ways     to     estimate     them
(Section~\ref{sec:models}).   The first  is the  widely  used diagonal
approximation, which has no  spatial or temporal correlation and whose
diagonal    elements    consist     of    sensor    noise    variances
(Section~\ref{sec:diagonal}).  The second model is a Kronecker product
approximation of a temporal  covariance and a spatial covariance under
the assumption that a temporal covariance and a spatial covariance are
independent   and    separable   \cite{huizenga2002}.    Parameterized
\cite{huizenga2002} and  unparameterized \cite{deMunck2002} variations
of this approximation  are available.  This paper deals  only with the
unparameterized  approximation  estimated  in the  maximum  likelihood
framework suggested in \cite{deMunck2002} (Section~\ref{sec:ML}).  Two
novel and  more complex  models that do  not employ the  assumption of
independence and  separability of temporal and  spatial covariance are
proposed          in          Section~\ref{sec:orthogonal}         and
Section~\ref{sec:independent}.   The  first   model  is  a  multi-pair
Kronecker product approximation based  on orthogonal spatial basis and
the second one is a variant based on independent spatial basis.

Since  the  goal  of   better  noise  characterization  by  covariance
modelling is to improve results  of inverse algorithms, we have chosen
to test and  compare all models described in this  paper by using them
in   algorithms   for  source   localization.    Performance  of   the
approximations in reconstructing dipole  locations and time courses is
tested  using a  large number  of  simulated single  dipole data  sets
constructed  from empirical  data for  background noise  and simulated
dipole  sources covering a  wide range  of locations  and orientations
(Section~\ref{sec:comparison}).

\section{Models and Methods}
\label{sec:models}
This section  describes all covariance models compared  in this paper.
Among  them,  two  novel   and  relatively  more  complex  models  are
introduced  and  explained.  Models  are  presented in  the  order  of
increasing complexity.

\subsection{Diagonal approximation}
\label{sec:diagonal}
It is common when solving the inverse problem to model covariance as a
diagonal  or even as  an identity  matrix.  In  the diagonal  case the
approximated   full   spatiotemporal   covariance  is   expressed   as
$\mathbf{COV} \propto \mathbf{T} \otimes \mathbf{S}$, where {\bf T} is
the temporal covariance which is taken  to be the identity; {\bf S} is
the diagonal  spatial covariance with  elements of the  diagonal being
sensor variances;  and $\otimes$ is the Kronecker  product. This model
is easy  to estimate.  It  has only $L$  parameters, where $L$  is the
number of sensors.  At the same  time this simple model is better than
not  using any covariance  estimate at  all as  in the  Ordinary Least
Squares (OLS) approach.

\subsection{Unparameterized Kronecker product model}
\label{sec:ML}
De  Munck et  al.~\cite{deMunck2002}  proposed a  more complex  model,
based on  the assumption that  temporal and spatial covariance  of the
background in MEG and EEG  are independent and separable.  This allows
one  to cast  spatiotemporal  covariance  in the  form  that uses  the
Kronecker product:
\begin{eqnarray}
  \mathbf{COV} \approx \mathbf{T} \otimes \mathbf{S} = \left [
    \begin{array}{cccc}
      \mathbf{t}_{11} \mathbf{S} & \mathbf{t}_{12} \mathbf{S} & \cdots 
      & \mathbf{t}_{1C} \mathbf{S} \\
      \mathbf{t}_{21} \mathbf{S} & \mathbf{t}_{22} \mathbf{S} & \cdots 
      & \mathbf{t}_{2C} \mathbf{S} \\
      \vdots & \vdots &  & \vdots \\
      \mathbf{t}_{C1} \mathbf{S} & \mathbf{t}_{C2} \mathbf{S} & \cdots 
      & \mathbf{t}_{CC} \mathbf{S} 
    \end{array}
  \right ]. \label{eqn:onepair}
\end{eqnarray}
Temporal  covariance is  a  $C\times  C$ matrix  {\bf  T} and  spatial
covariance is an  $L\times L$ matrix {\bf S}, where  $C$ is the number
of   time    samples   and   $L$    is   the   number    of   sensors.
In~\cite{deMunck2002}  the  temporal  covariance  matrix  {\bf  T}  is
normalized to  one. Both matrices  are unparameterized.  Note,  that a
single set  of variances needs to  be divided between  the spatial and
temporal  matrices and  that normalization  of  {\bf T}  takes away  a
degree of freedom. In this case  the number of parameters that need to
be  estimated  equals  $L(L+1)/2+C(C-1)/2$.   This is  less  than  all
elements  of   both  matrices  because  both   are  symmetric.   These
parameters are estimated from the data using a Maximum Likelihood (ML)
method  as  described  in~\cite{deMunck2002}.   In deriving  these  ML
estimators,  the  single  trial  background  data was  assumed  to  be
Gaussian  with a covariance  that factored  into separate  spatial and
temporal matrices.   This Gaussian distribution was  used to construct
the  likelihood  of  the  single  trial background  data,  from  which
estimators were derived for the parameters of the spatial and temporal
covariance  matrices that  maximized this  likelihood.   The resulting
estimators are a coupled set of equations between spatial and temporal
parameters that may be solved using an iterative technique.

\subsection{Orthogonal basis multi-pair model}
\label{sec:orthogonal}
In the  Kronecker product model  above, all spatial components  of the
background have the same temporal covariance structure. This statement
describes  the main  assumption of  the single-pair  Kronecker product
model.   In  order to  motivate  multi-pair  models,  we re-write  the
single-pair  Kronecker product  model  in the  following way.   First,
perform a spectral decomposition of the spatial covariance:
\begin{equation}
  \mathbf{T} \otimes \sum_{l = 1}^{L} \sigma^{l} \bm{{\mathcal S}}^l,
  \label{eq:spectral_onepair}
\end{equation}
where  $\bm{{\mathcal  S}}^l =  v_l  v_l^T$  is  an orthonormal  basis
component represented as  a singular matrix.  This form  is one of the
conventional ways  of writing a  spectral representation of  a matrix.
Using   the   identity   $\mathbf{A}\otimes(\mathbf{B}+\mathbf{C})   =
\mathbf{A}\otimes\mathbf{B}    +   \mathbf{A}\otimes\mathbf{C}$,   the
expression~(\ref{eq:spectral_onepair}) can be represented as:
\begin{equation}
  \sum_{l = 1}^{L} \mathbf{T} \otimes \sigma^{l} \bm{{\mathcal S}}^l = 
  \sum_{l = 1}^{L} \sigma^{l} \mathbf{T} \otimes \bm{{\mathcal S}}^{l}
  \label{eq:proto_multipair}
\end{equation}
The left hand side of the equation~(\ref{eq:proto_multipair}) makes it
obvious that each spatial  component has the same temporal covariance.
The  contribution  of each  temporal  covariance  is  weighted by  the
variance of  the corresponding orthonormal spatial  component, as seen
from the right hand  side of~(\ref{eq:proto_multipair}).  In the final
sum such  weighting of temporal covariances makes  no difference since
they all are the same.

In  a more  realistic case  it is  easy to  picture a  situation where
several  noise  generators   having  distinct  spatial  patterns  (or,
similarly,  belonging  to   separate  spatial  components)  also  have
different and independent temporal  structures.  They can have a focal
or  a  distributed  spatial  pattern  that can  be  described  by  the
corresponding spatial component.  Furthermore, some background sources
of noise that are external to  the head origin can also be captured by
such a model.

We propose an alternative multi-pair model of the spatiotemporal noise
covariance  as  a  {\it  series}  of  orthonormal  spatial  components
$\bm{{\mathcal S}}^l$  of the background data  and their corresponding
temporal covariance matrices $\bm{{\mathcal T}}^l$ expressed as:
\begin{eqnarray}
  \mathbf{COV} \approx \sum_{l = 1}^{L} \bm{{\mathcal T}}^l \otimes
  \bm{{\mathcal S}}^l .\label{eq:series_form}
\end{eqnarray}
The model is build on the following assumptions:
\begin{enumerate}
\item[A1]   Spatiotemporal  noise  is   generated  by   $L$  spatially
  orthogonal generators which do  not change their location during the
  period of interest.
\item[A2] Each  spatial component has  a time course  independent from
  those of other components.
\item[A3] Superposition of time  courses of each component measured at
  the    sensors    has   a    Gaussian    distribution   with    zero
  mean. \label{itGaussian}
\end{enumerate}

Let's  assume for  now that  we are  given the  orthonormal components
$\bm{{\mathcal S}}^l$. (We will discuss a way to obtain those later in
the section.)  As in~(\ref{eq:spectral_onepair}) the orthonormal basis
spans the whole sensors space  having dimension $L$.  The next task is
to  estimate  temporal   covariances  $\bm{{\mathcal  T}}^l$  of  each
component.   This  can be  done  due to  the  assumption  A3 that  the
background is Gaussian and  using the Maximal Likelihood estimation as
demonstrated in  the Appendix~\ref{sec:multi_pair_derivation}.  Assume
that  we have  $M$ single  trial noise  samples $\mathbf{E}_{m}$  -- a
$L\times C$ matrix  with $L$ number of sensors and  $C$ number of time
points. Now we can summarize  orthogonal basis multi-pair model in the
following way:
\begin{eqnarray}\nonumber
  \mathbf{COV} \approx
  \widetilde{\mathbf{COV}} &\equiv& \sum_{l=1}^L  \bm{{\mathcal T}}^l
  \otimes \bm{{\mathcal S}}^l, \\\nonumber
  \bm{{\mathcal T}}^{l} &=& \frac 1{M^{2}} \sum_{m=1}^M 
  \mathbf{E}_{m}^{T} \bm{{\mathcal S}}^l \mathbf{E}_{m},\\
  \bm{{\mathcal S}}^l &=& v_l v^T_l.
  \label{eq:svd_summary}
\end{eqnarray}
Expression~(\ref{eq:svd_summary}) represents a  model for the averaged
noise hence  the normalizing factor in the  estimate of $\bm{{\mathcal
    T}}^{l}$  is squared. Note  that although  the form  of expression
(\ref{eq:series_form})  may suggest  that  it is  a  summation of  $L$
single pair models, this  is not so.  Components $\bm{\mathcal S}^{l}$
are  not covariances and  furthermore they  are singular  because they
represent a dimension in an $L$ dimensional space.

Another important  feature expected of  a covariance model to  make it
useful in  the analysis is  a computationally manageable  inverse. The
inversion of this  multi-pair model can be expressed  by (see Appendix
\ref{sec:multi_pair_inversion} for details)
\begin{eqnarray}
  \widetilde{\mathbf{COV}}^{-1} &=& \sum_{l=1}^L 
  (\bm{{\mathcal T}}^l)^{-1} \otimes \bm{{\mathcal S}}^l.
  \label{eq:ortho_inv}
\end{eqnarray}
This inversion is easily  manageable. Its running time is $O(LC^{3})$,
similar to that for the one-pair models.

While  this  model  has  far  fewer  free  parameters  than  the  full
spatiotemporal covariance estimate (\ref{eqn:smple_cov}) the number of
parameters is larger  than that of models described  above (though not
by a  great amount). The $L$ spatial  components $\bm{{\mathcal S}}^l$
consist of  $L(L-1)/2$ parameters plus $LC(C+1)/2$  parameters for all
temporal covariances. The increase  in parameters provides an increase
of  expressive power that  is greater  than of  the single  pair model
(\ref{eqn:onepair}).   Furthermore,  it is  free  from assumptions  of
identical character of the noise  over all spatial locations and hence
may capture more information about structure of the noise.

We  derived  this model  without  any  assumptions  about how  spatial
orthogonal components were obtained.  That means that in principle any
set of such components should  keep the validity of the derivation and
retain the invertibility  of the model. However, since  we assume that
the background has  the distribution close to the  Gaussian then it is
natural to use singular value decomposition (SVD) to obtain orthogonal
spatial components.  In this work we  use SVD to  estimate the spatial
orthogonal components for the  orthogonal basis multi-pair model.  SVD
of the noise  data collected in a matrix  $\mathbf{A}$ by stacking $M$
single   trial  spatiotemporal   samples  $\mathbf{E}_m$   looks  like
$\mathbf{A}  = \mathbf{U}\mathbf{\Sigma}\mathbf{V}^{T}$.  $\mathbf{U}$
is a $CM\times L$  orthogonal matrix; $\mathbf{\Sigma}$ is an $L\times
L$ diagonal matrix with  singular values of $\mathbf{A}$ $\lambda_{l}$
as diagonal  elements; and $\mathbf{V}$  is an $L\times  L$ orthogonal
matrix  of spatial  components.   Each row  of  $\mathbf{V}^{T}$ is  a
spatial  component  $v_l$ that  is  used  to  form orthogonal  spatial
components $\bm{{\mathcal S}}^l$ of our model~(\ref{eq:series_form}).

Very often SVD is used for dimensionality reduction when only the most
significant singular values are accounted for and all other values are
neglected~\cite{LKS05}.   In this application  we do  not use  SVD for
this  purpose.  Dimensionality reduction,  i.e.  approximation  of the
full covariance  is already performed  in~(\ref{eq:series_form}) based
on the  stated assumptions. The sole  purpose of SVD  in estimation of
this  model  is  finding   spatial  orthogonal  components  and  their
corresponding time courses.

\subsection{Independent basis multi-pair model}
\label{sec:independent}
The model  introduced in the  previous section has  greater expressive
power  than  the   models  presented  previously.   Nevertheless,  its
requirement of  orthogonality may  be inappropriate for  MEG/EEG data.
In  the  general  case noise  sources  do  not  have to  be  spatially
orthogonal to one another. In  an effort to remove this restriction we
suggest a generalization, that does not restrict spatial components to
be orthogonal. Instead we consider independent spatial components.

Let us assume that we  are given independent spatial components of the
background.    Denote  each  independent   component  as   $w_l$,  its
corresponding spatial matrix as  $\bm{{\mathcal R}}^l = w_l w^T_l$ and
the  matrix, in  which  the $l^{th}$  row  is the  $w_{l}$ vector,  as
$\mathbf{W}^{-1}$.  In  this framework  the only assumption  needed is
independence  of  each  $w_{l}$  from  the  rest  of  the  components.
Estimation  of  temporal   covariances  of  each  independent  spatial
component can  be performed in  the Maximum Likelihood  framework (see
Appendix~\ref{sec:multi_pair_derivation}).
\begin{eqnarray}\nonumber
  \mathbf{COV} \approx  
  \widetilde{\mathbf{COV}}  &\equiv& \sum_{l=1}^L \bm{{\mathcal T}}^{l}
  \otimes 
  \bm{{\mathcal R}}^{l} \\\nonumber
  \bm{{\mathcal T}}^{l} &=& \frac 1{M^{2}} \sum_{m=1}^M \mathbf{E}_{m}^{T} 
  \mathbf{W}\mathbf{W}^{T} \bm{{\mathcal R}}^l \mathbf{W}\mathbf{W}^{T}
  \mathbf{E}_{m} \\
  \bm{{\mathcal R}}^l &=&  w_l w^T_l.
  \label{eqn:ica_form}
\end{eqnarray}
Here $w_l$  stands for the  $l$-th row vector of  $\mathbf{W}^{-1}$ as
described above and $M^{2}$ is due to the averaged noise modeling.

The model is  more general compared to the  one using orthogonal bases
from Section  \ref{sec:orthogonal}. It can be constructed  in terms of
any spatially  independent basis  set, which also  includes orthogonal
sets.  The  number of  free parameters of  the model has  increased to
$L^{2}+LC(C-1)/2$. At the same  time its inversion is still manageable
and can  be expressed as  (see Appendix \ref{sec:multi_pair_inversion}
for details)
\begin{equation}
  \widetilde{\mathbf{COV}}^{-1} = \sum_{l=1}^L (\bm{{\mathcal
      T}}^{l})^{-1} \otimes[ \mathbf{W} \mathbf{W}^T \bm{{\mathcal
      R}}^l \mathbf{W} \mathbf{W}^T].
  \label{eq:ica_inv}
\end{equation}
This  makes   the  model  useful   in  the  analysis.    However,  the
computational  cost of  this operation  is greater  than that  for the
orthogonal     basis     multi-pair      model.      It     is     now
$O(L(C^{3}+2L^{3})+L^{3})$,  where   the  $2L^{3}$  part   comes  from
repeated  multiplication   of  independent  components  $\bm{{\mathcal
    R}}^l$   by  the   combination  $\mathbf{W}   \mathbf{W}^T$.   The
additional   $L^{3}$  component   is  the   time  to   calculate  this
combination.  Nevertheless in localization algorithms, like in the one
we used  in this study, the  covariance matrix does  not change across
iterations  and its  inverse  needs  to be  computed  only once.   The
increase in computation does not significantly influence total running
time of  inverse procedures and  thus represents a small  tradeoff for
the generality of the model over the one using an orthogonal basis.

The  above reasoning  applies  to any  independent  components of  the
measured signal.  It is crucial  to adopt an optimal way of estimating
such  components  (with  respect  to  some  criterion).   A  class  of
algorithms  that  uses an  optimality  criterion  to find  independent
components arises in  the field of blind source  separation: the class
of Independent Component Analysis  (ICA) algorithms.  We adopt such an
algorithm  for this  work as  a tool  to discover  independent spatial
components.

Different     ICA     algorithms      have     been     applied     to
MEG~\cite{VSJ+00,TPM+02b,TPM+02,ZMN+00,CaoJ2000b}                   and
EEG~\cite{MJB+97,MWJ+99,JMH+00} data. Among  them the most widely used
are  second order blind  identification (SOBI)  \cite{BACM93}, Infomax
\cite{Bell95}, and fICA \cite{Hyvarinen+97}. Our goal in this paper is
not  to  evaluate  performance  of  different ICA  algorithms  in  the
suggested framework; that is work for a subsequent study. Rather, here
we   chose  one   algorithm   to  demonstrate   the  feasibility   and
applicability  of  ICA  for  finding  independent  spatial  components
$\bm{{\mathcal  R}}^l$ for modeling  background noise.   A preliminary
study  compared  Infomax   and  SOBI  algorithms.   Initial  estimates
suggested that  SOBI gave a  better approximation to the  structure of
the full spatiotemporal covariance.  In this paper we have used a SOBI
algorithm implementation provided by ICALAB toolbox~\cite{ICALAB} with
the  default   settings.   The  SOBI   algorithm  was  also   used  in
\cite{TPM+02,TPM+02b} and  the motivation for applying it  to MEG data
is given there.

Assume that we have $M$ times  single trial noise data and that an ICA
algorithm   is  applied  to   this  data   constructed  in   a  martix
$\mathbf{A}$:
\begin{equation}
  \left ( 
    \begin{array}{c}
      \mathbf{E}^T_1 \\
      \vdots \\
      \mathbf{E}^T_M
    \end{array}
  \right ) = 
  \mathbf{A} = 
  \left(
    \begin{array}{c}
      \mathbf{U}_1 \\
      \vdots \\
      \mathbf{U}_M
    \end{array}
  \right ) \mathbf{W}^{-1}.
  \label{eqn:ica_mixing}
\end{equation}
In  the above  expression $\mathbf{W}$  is  an $L  \times L$  unmixing
matrix obtained  as a result  of an ICA algorithm,  the $\mathbf{U}_m$
form  a $CM\times  L$ matrix  of independent  components $\mathbf{U}$.
This  $\mathbf{W}$ matrix is  the matrix  used in~(\ref{eqn:ica_form})
and~(\ref{eq:ica_inv}).

Although ICA is widely used for dimensionality reduction, we note that
in this application no  dimensionality reduction is performed based on
ICA; the technique is used exclusively as a discovery tool.

\section{Comparing performance of covariance models}
\label{sec:comparison}
Many different approaches  can be used to evaluate  how well different
models  approximate the  full spatiotemporal  covariance.  Calculating
some  norm   of  the  difference  between  true   covariance  and  its
approximating model  is a common  measure, for instance  the Frobenius
norm.   Kullback-Leibler  divergence \cite{KL51a}  is  also a  popular
measure. A scatterplot  can be a good "visual"  tool for comparison as
described  for  example  in \cite{huizenga2002}.   However,  different
measures  can  give  different  and even  contradictory  results  when
comparing  several  models.   The   ultimate  goal  of  modeling  full
spatiotemporal  covariance is to  achieve better  inverse performance.
Thus,  evaluation of  model  performance in  an  inverse algorithm  is
adopted as the main comparison tool  in this study.  A large number of
single dipole data sets were analyzed using different background noise
models.  Each data set was  constructed using empirical whole head MEG
background data together with simulated dipole sources.

\begin{figure}[ht]
  \centering \includegraphics[width = 3.45in]{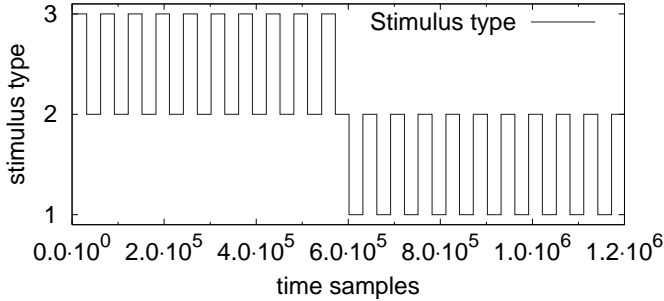}
  \caption{The manner in which stimuli were applied in the experiment.
    x-axis shows time samples and y-axis - stimulus types, where type
    1 - right hand, 2 - no stimulus, 3 - left hand.}
  \label{fig:data_setup}
\end{figure}

The empirical MEG data and anatomical information used for performance
comparisons were acquired in the following experiment:

Electrical median nerve stimulation  at the motor twitch threshold was
applied using  a block design  of 30s  on, 30s off  for a total  of 10
blocks for each  of 8 runs. The stimulus  alternated across runs, with
four runs each of left side stimulation and of right side stimulation.
The ISI (interstimulus interval) was randomized between 0.25 and 0.75s
(Fig. \ref{fig:data_setup}).  Since there  is no stim for long periods
this design provides  a large sample of brain  noise data. Which might
be  useful  in other  noise  studies.  Data  were  collected  on a  4D
Neuroimaging  Neuromag-122  whole-head  gradiometer  system  with  122
channels \cite{Ahonen93}. The experiment  used a male subject, age 38,
sampling rate was set to 1000Hz. In this paper data from sensor 51 was
not used  since its  output had too  many artifacts, leaving  only 121
useable channels.

This somatosensory data set contained high frequency transient signals
resulting from the electrical stimulus. These transients are distorted
when  filtered  with a  linear  filter,  due  to ringing  effects  and
adversely affect nearby data, including the expected early response at
about 20ms post-stimulus.   To avoid this and still  be able to remove
low frequency drifts, a median filter was used as follows.  First, the
signal was filtered with a median  filter of window size set to obtain
1Hz low  pass filter  (1000 samples in  our case). Second,  the result
from  the previous step  was subtracted  from initial  measurements to
obtain  1Hz  high-pass  filtered  signal.  The large  60Hz  noise  and
harmonics in  the data were reduced  but not removed  by replacing the
points in the power spectrum in  the data near 60Hz and harmonics with
values that interpolated between adjacent power spectrum points.

Noise samples of 35ms latency were extracted from the prestimulus area
of right  side stimulation  no earlier than  300ms after  the previous
stimulus was  applied. Since filtering can  affect temporal covariance
it  is  important to  estimate  covariance  after  filtering has  been
applied. All  covariance models in  this section were  estimated using
this  continuous  background  data,  scaled  appropriately  to  obtain
covariances of the noise averaged over 602 samples. All the covariance
models  yielded  similar   sensor  variances,  differences  were  only
observed in the correlation structure.

Continuous  noise samples  were combined  in  a way  that allowed  for
averaging  over 602  independent samples.  This approach  supplied six
different average  noise data sets.  Different signal to  noise ratios
(SNR) for single dipole problems  were obtained by scaling these noise
samples before combining them with  simulated data. The measure of SNR
used  in this  paper was  constructed by  squaring all  values  of the
signal  vector and  then adding  them together  (inner product  of the
signal vector) and dividing this value by squared and summed values of
the noise vector; a square root was taken from the obtained ratio.

\begin{figure}[ht]
  \centering \includegraphics[width = 5.3in]{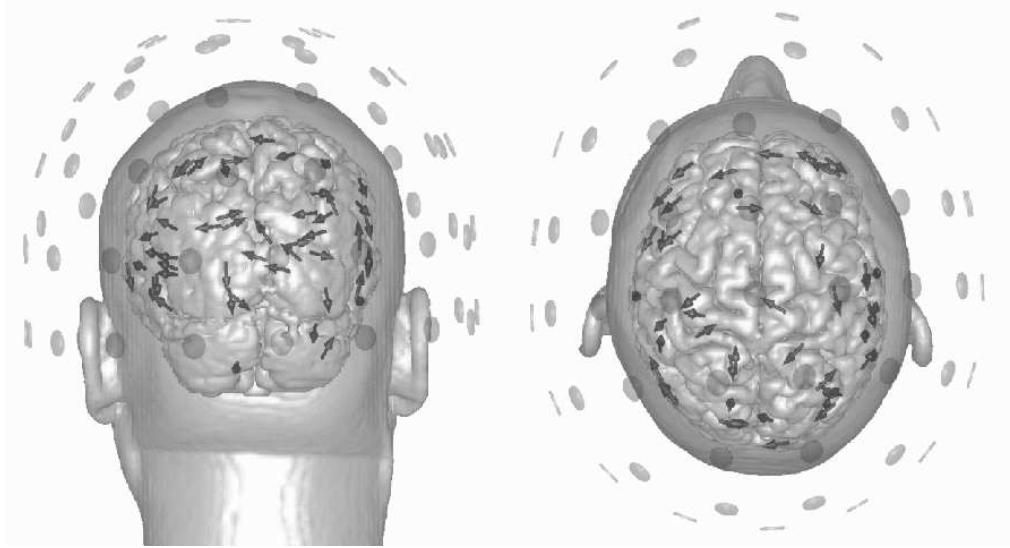}
  \caption{Dipoles randomly scattered over the cortex. These dipoles
    give fairly uniform coverage of the cortex and with different
    orientations create many possible realistic sources.}
  \label{fig:dipoles}
\end{figure}

The locations and  orientations of fifty dipoles were  drawn at random
from the  gray matter voxels that  had been tagged  from the subject's
anatomical  MRI used  in the  empirical MEG  experiment.  In  order to
mitigate the  effect of depth  vs. strength that would  complicate the
interpretation of  results, the area from which  random locations were
drawn was  constrained to  be further than  five centimeters  from the
head center (Fig. \ref{fig:dipoles}).

Each single dipole  problem consisting of 121 sensor  values over 35ms
(121x35 matrix) was constructed in the following way:
\begin{enumerate}
\item For  each dipole in the set  of 50 a sinusoidal  time course was
  used. 
\item  This dipole  was projected  to the  sensor space  using Sarvas'
  spherical head model \cite{Sar87}.
\item To  the simulated signal  one of the  six noise sample  sets was
  added.  Noise  was appropriately scaled in advance  according to the
  intended SNR.
\end{enumerate}

\begin{figure}[ht]
  \begin{center}
    \begin{tabular}{c|rrr}
      \raisebox{0.3in}{\rotateleft{\small{\tt instance 1}}} &
      \includegraphics[width =
      1.66in]{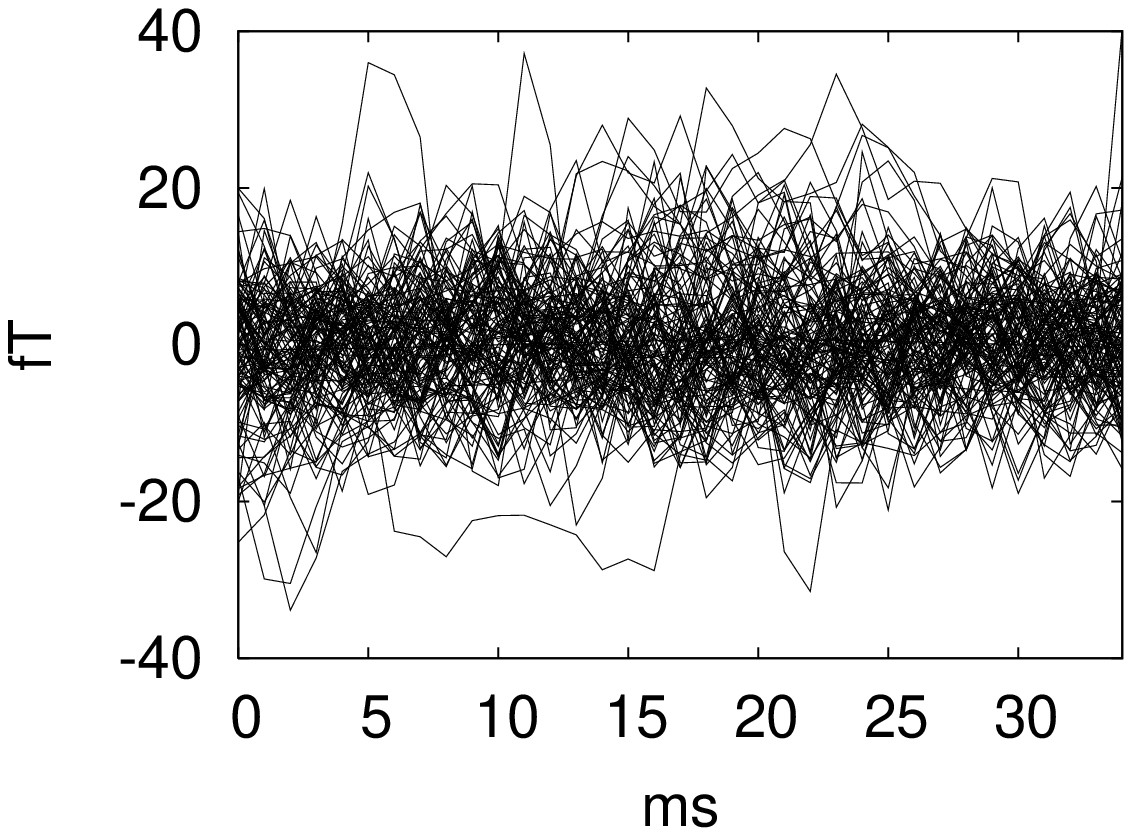} &
      \includegraphics[width =
      1.66in]{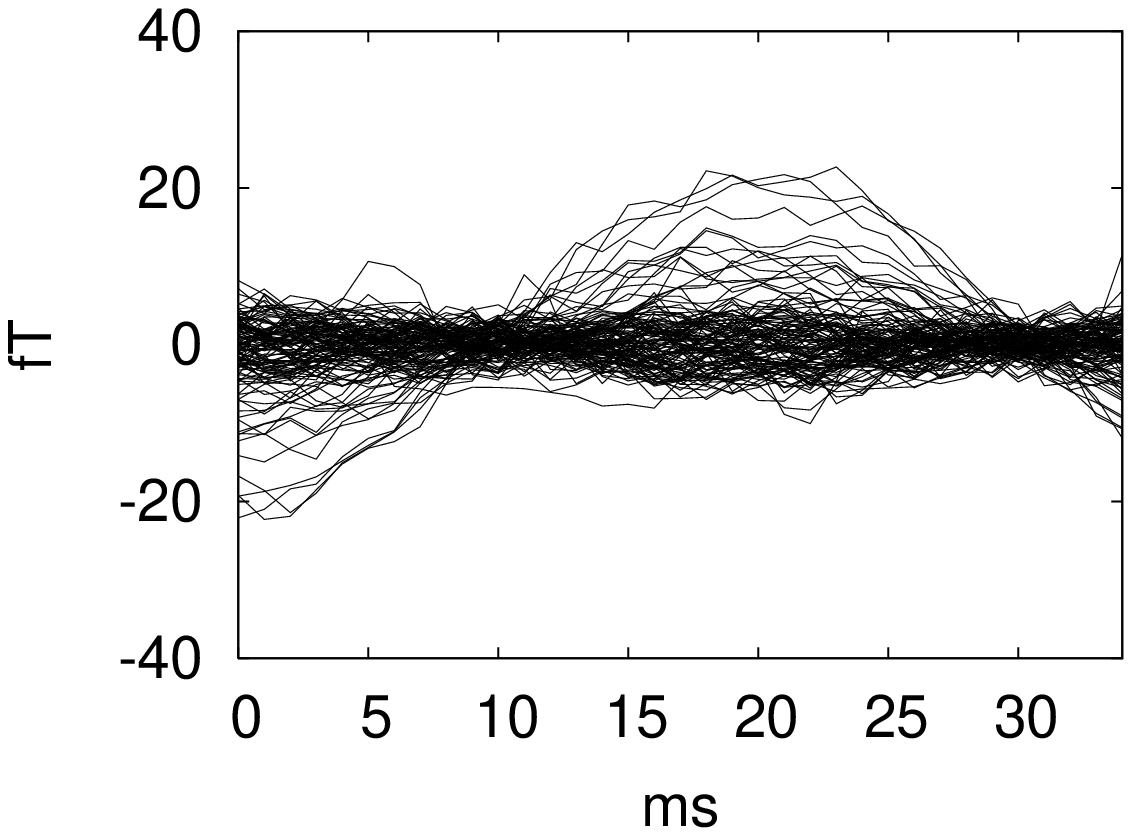} &
      \includegraphics[width =
      1.66in]{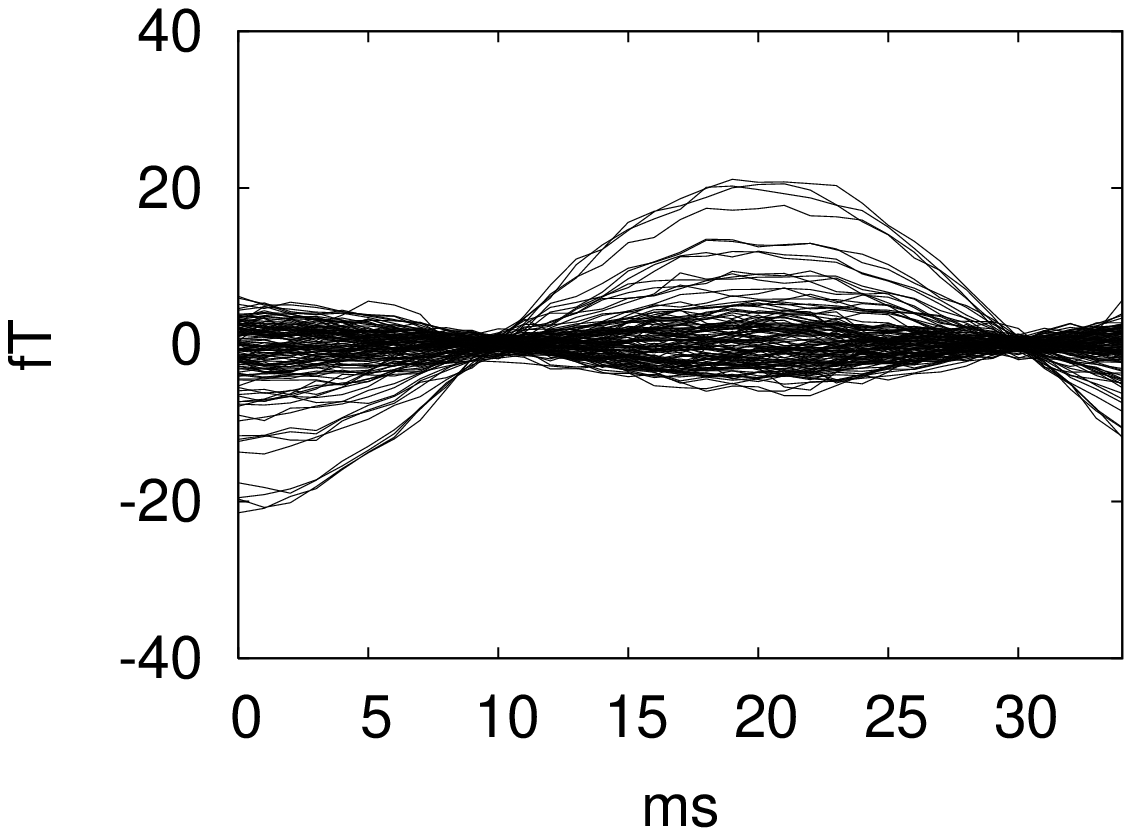} \\
      \raisebox{0.3in}{\rotateleft{\small{\tt instance 2}}} &
      \includegraphics[width =
      1.66in]{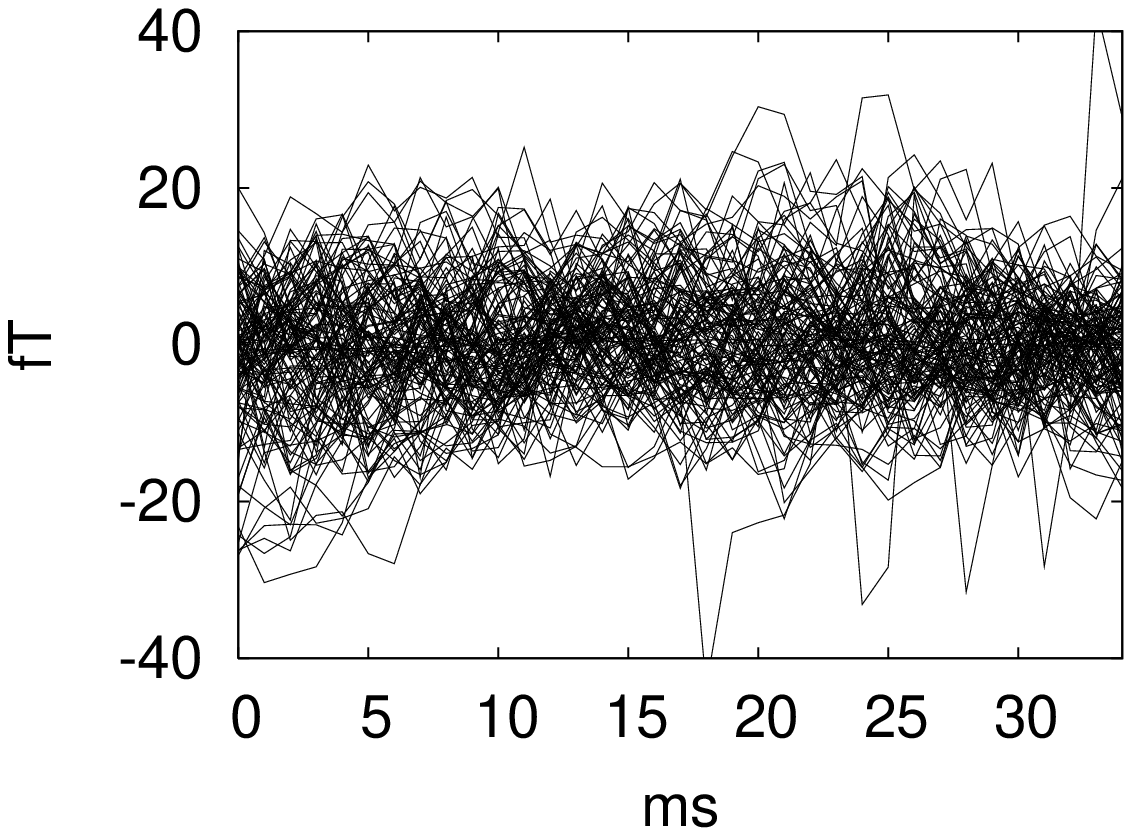} &
      \includegraphics[width =
      1.66in]{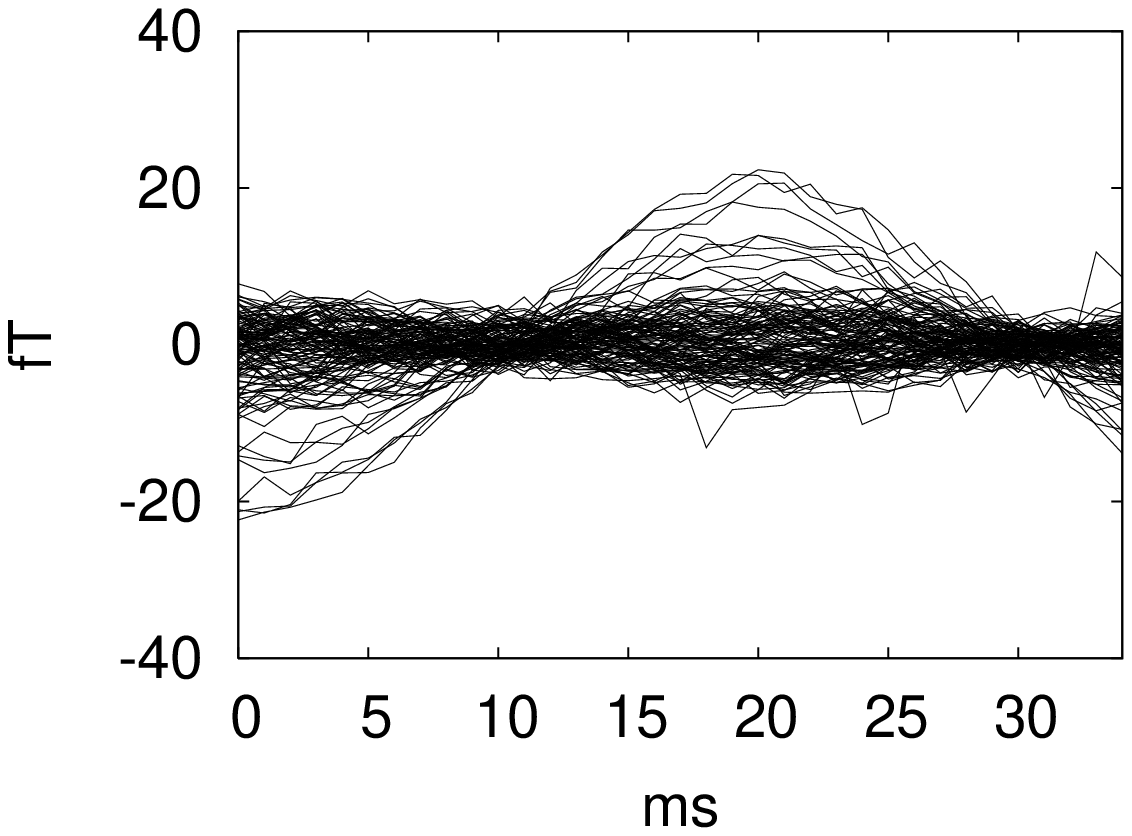} &
      \includegraphics[width =
      1.66in]{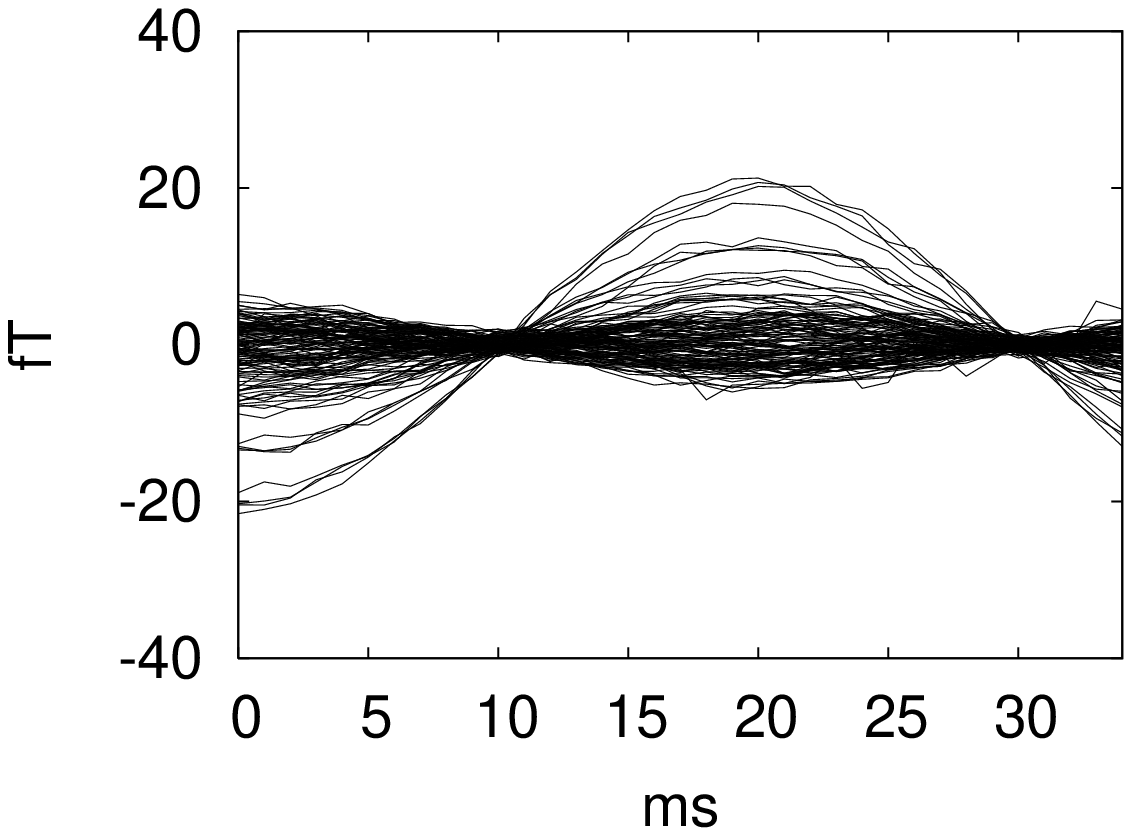} \\
      \raisebox{0.3in}{\rotateleft{\small{\tt instance 3}}}&
      \includegraphics[width =
      1.66in]{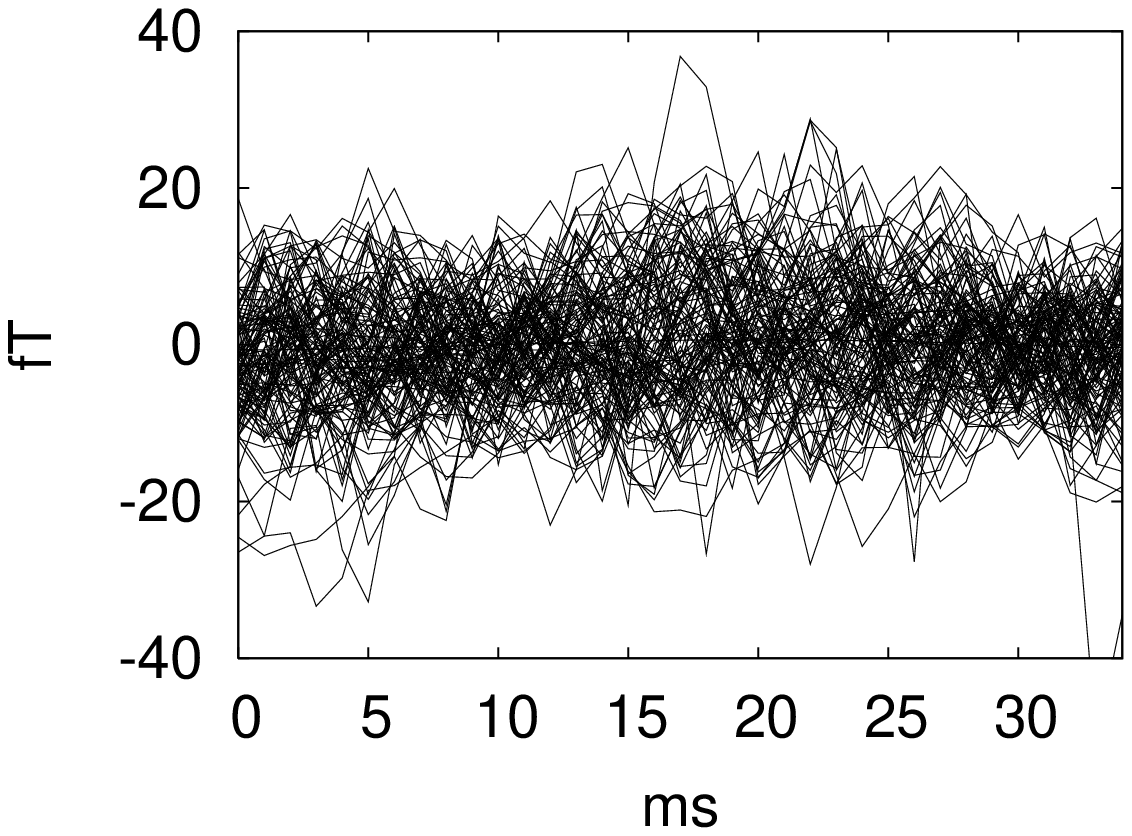} &
      \includegraphics[width =
      1.66in]{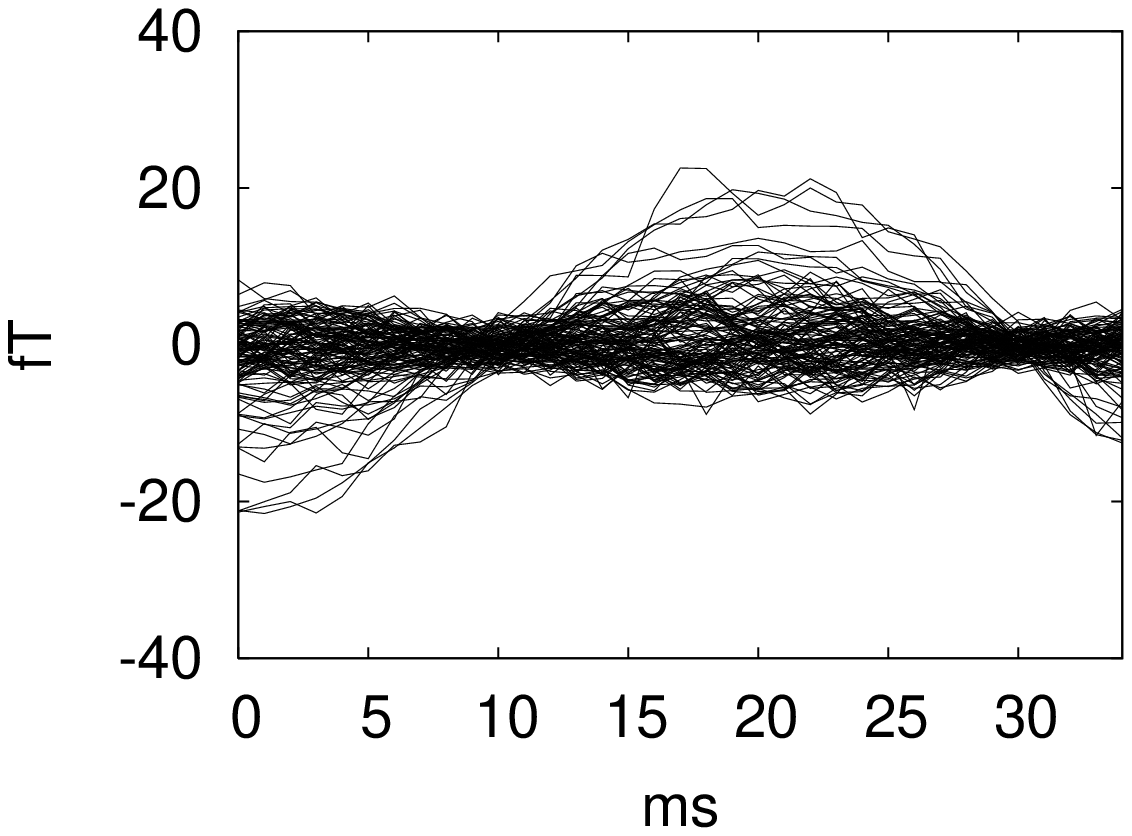} &
      \includegraphics[width =
      1.66in]{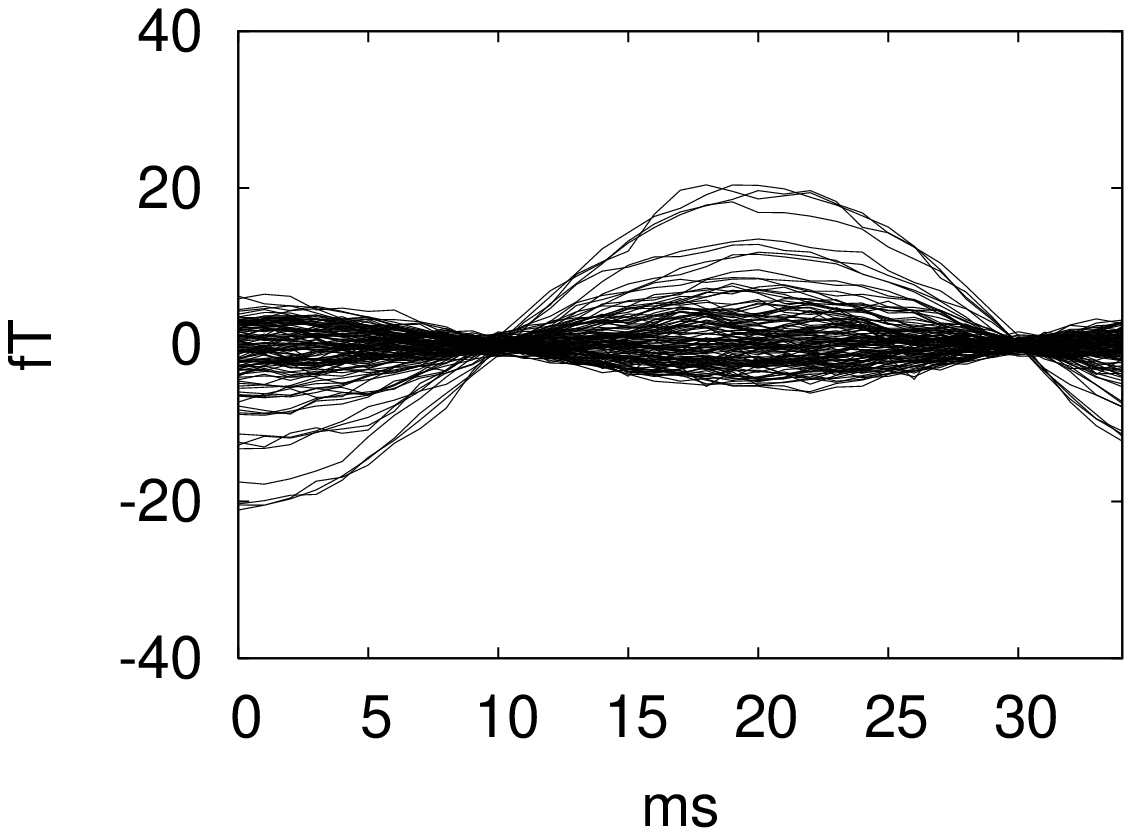} \\
      \cline{2-4}
      \multicolumn{2}{r}{\small{\tt SNR 0.3}}& \small{\tt SNR 1.0} &
      \small{\tt SNR 3.0} \\
    \end{tabular}
  \end{center}
  \caption{Three noise sample sets and  three signal to noise ratios for
    one of the dipoles. Columns have  the same signal to noise ratio and
    rows have the same noise instance added to the simulated dipole.}
  \label{fig:overplots}
\end{figure}

The total  number of  single dipole problems  run through  the inverse
solution  routine was  2400. This  number combines  50 dipoles  with 6
noise      kinds      for     each      and      8     SNR      values
(0.3,0.4,0.5,0.7,1.0,1.5,2.0,3.0).      Figure     \ref{fig:overplots}
demonstrates how  different the noise samples were  and what diversity
was introduced by changing SNR.

For  each data  set we  estimated the  location, orientation  and time
course of  a single  dipole that maximized  the likelihood,  using the
different background covariance models.   In these trials the location
and  orientation  of the  dipole  did  not  vary over  time.   Initial
attempts to solve for these parameters using an optimization algorithm
were  plagued by  local minima  problems,  which is  common in  dipole
inverse  algorithms  \cite{HMAC+98MSST,MRIVIEW-2002}.  The  degree  of
these  local minima  errors confounded  the errors  between background
covariance models.   To mitigate this confound we  employed a sampling
algorithm     using     Markov     Chain    Monte     Carlo     (MCMC)
\cite{Gelman95,JUN-ETAL-2005B,SCHMIDT-ETAL-1999A},  which  sampled the
location  and orientation  parameters  from the  likelihood using  the
maximum likelihood time course values  for a given set of location and
orientation  parameters.  From  these samples  we then  calculated the
mean values  of the parameters  and used this  as our estimate  of the
maximum likelihood result.  This reduced the local minima problems but
did not eliminate them.  To further reduce the local minima effects we
ran multiple  MCMC samplings for each  data set and  chose the results
that had  the highest likelihood.   We are confident that  the results
from this  set of procedures  primarily reflect the  errors associated
with the different background covariance models.

\begin{figure}[ht]
  \centering \includegraphics[width = 5in]{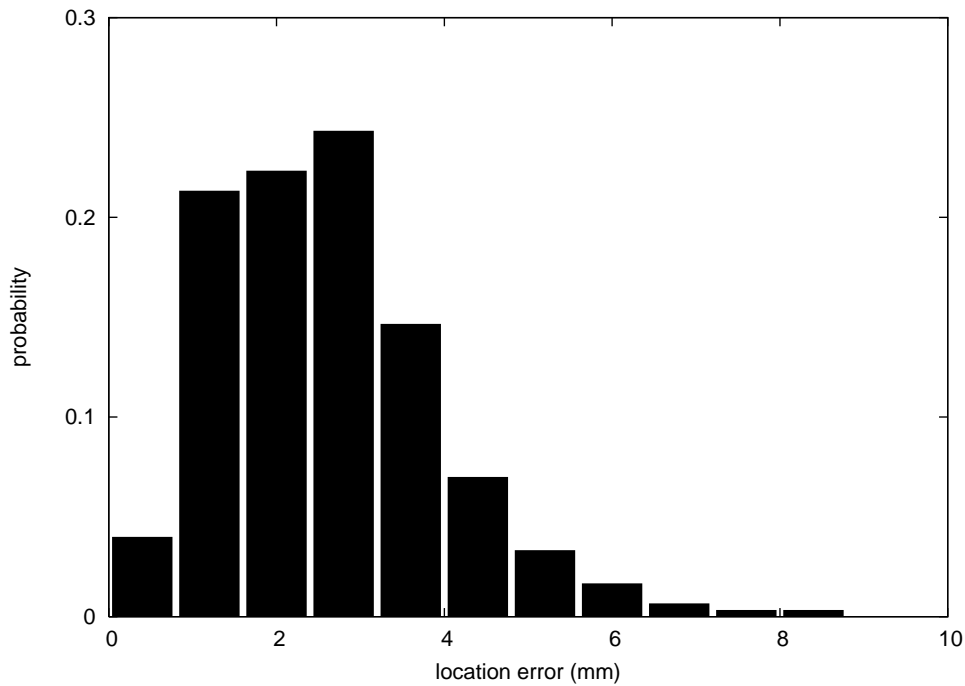}
  \caption{Location error histogram for the case of using the diagonal
    covariance approximation at SNR 1.0}
  \label{fig:hist}
\end{figure}

\begin{figure}[ht]\
  \subfigure[location errors]{
    \centering \includegraphics[height =
    2.5in]{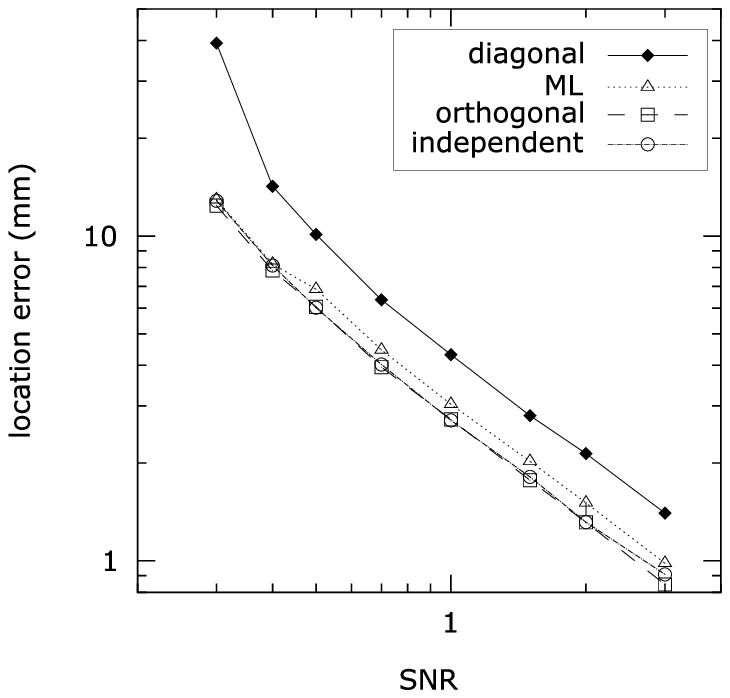}
    \label{fig:error_a}
  } \hspace{-0cm}
  \subfigure[time course errors]{
    \centering \includegraphics[height = 2.5in]{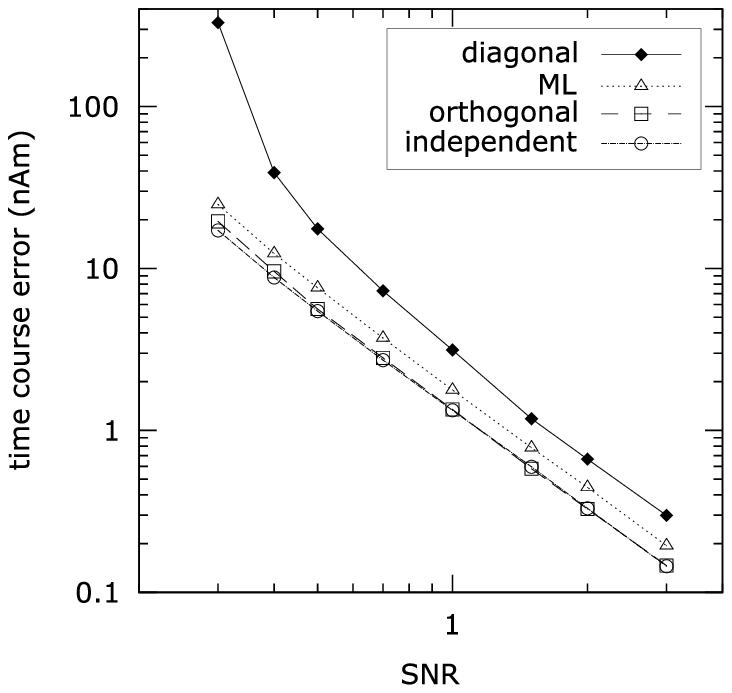}
    \label{fig:error_b}
  }
  \caption{Combined  90\% location  error measure~\subref{fig:error_a}
    and  combined 90\% time  course errors~\subref{fig:error_b}  for a
    series of single dipole problems.}
  \label{fig:error}
\end{figure}

There  are   300  results   for  each  SNR   value  for   each  model.
Figure~\ref{fig:hist}  shows a  histogram of  location errors  for the
diagonal  model with  SNR  of  1.0.  The  shape  of this  distribution
(non-Gaussian with large tails), is  typical for all of the models and
SNR values.  From each of  these distributions we calculated the point
on  the  error  axis below  which  90\%  of  the probability  mass  is
concentrated.  Figures  \ref{fig:error_a}  and \ref{fig:error_b}  show
these 90\% error plots for location  and time courses as a function of
SNR for each background covariance model.  Here the location error was
calculated as  the root mean squared  error and the  time course error
was calculated  as the root of  the squared error  averaged over time.
The two multi-pair  models had the lowest errors,  followed by the two
single pair models with  slightly higher errors. Finally, the diagonal
model had by far the largest errors.

\section{Discussion}

\begin{figure}[ht]
  \begin{center}
    \begin{tabular}{l|cccc|}\cline{2-5}
      \raisebox{0.71in}{\tiny{\tt 1}}&
      \includegraphics[height = 1.25in, width = 1.25in]{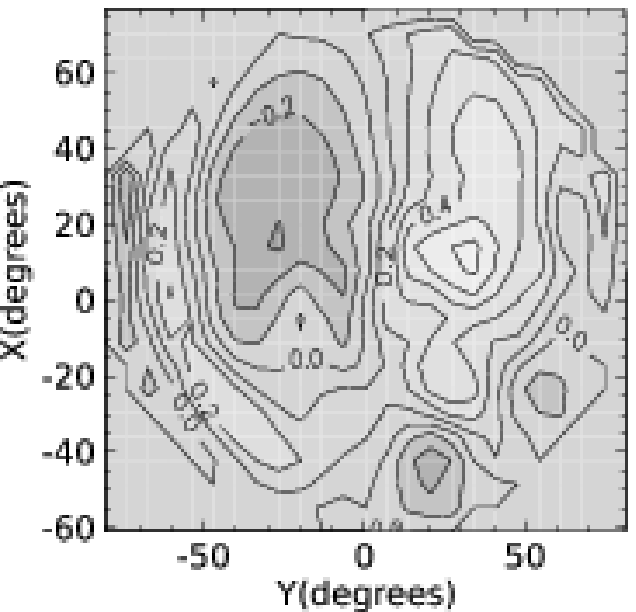}&
      \includegraphics[height = 1.25in]{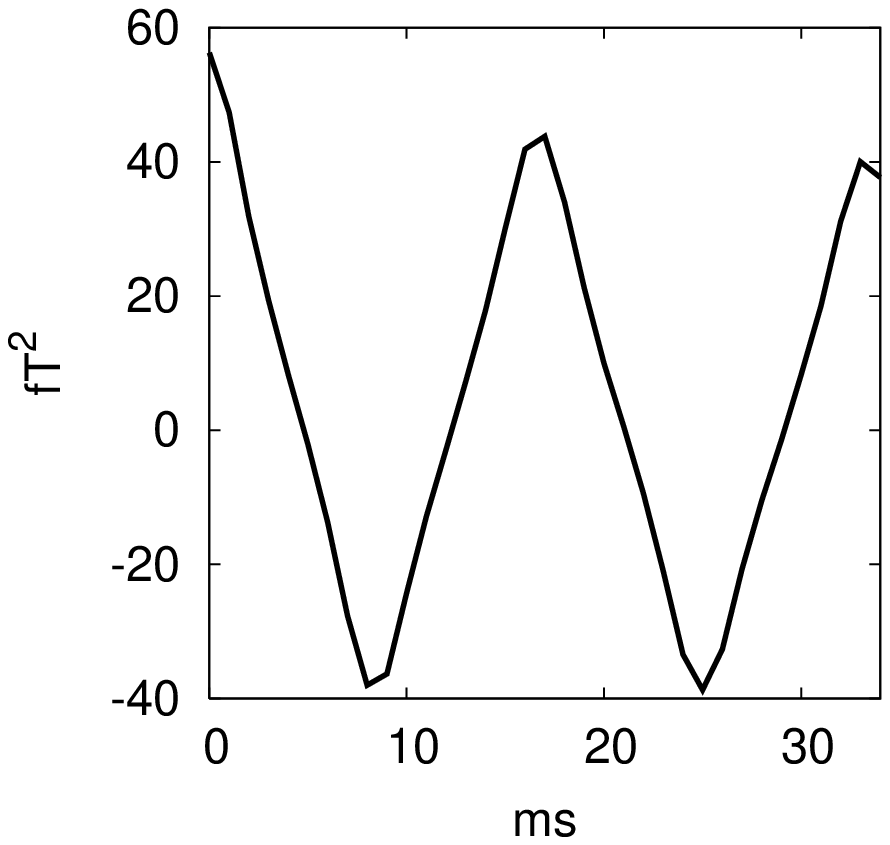} &
      \includegraphics[height = 1.25in, width = 1.25in]{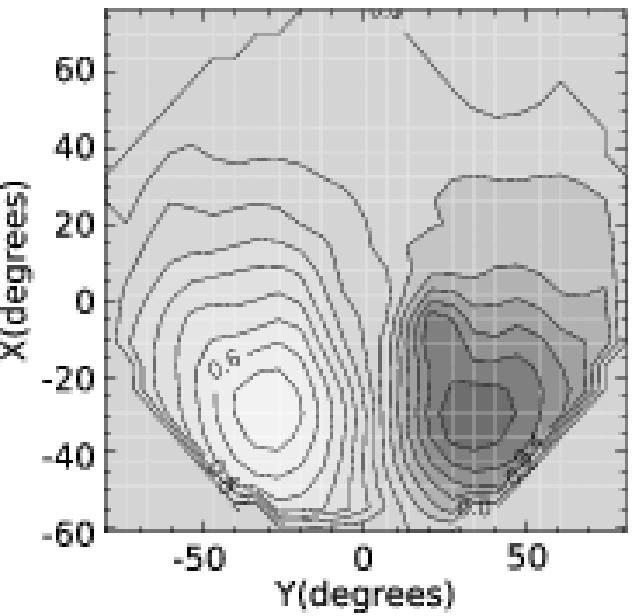} &
      \includegraphics[height = 1.25in]{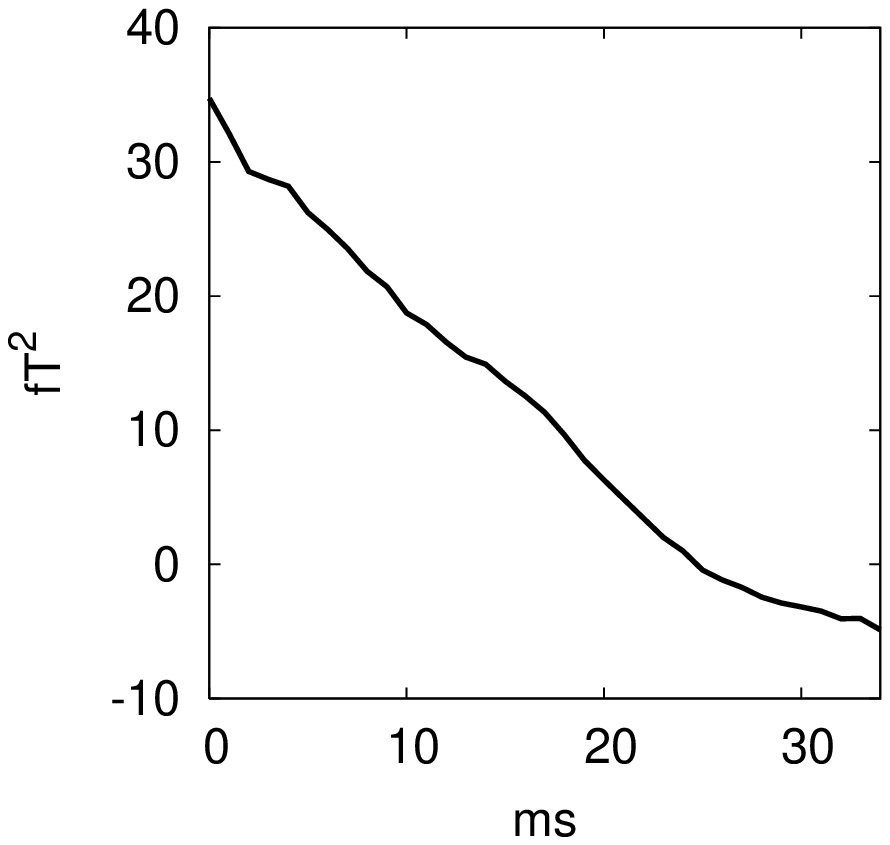}\\

      \raisebox{0.71in}{\tiny{\tt 2}}&
      \includegraphics[height = 1.25in, width = 1.25in]{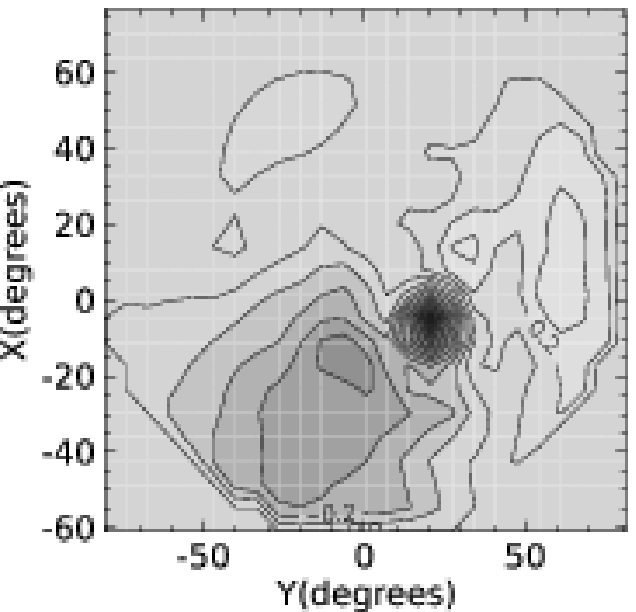} &
      \includegraphics[height = 1.25in]{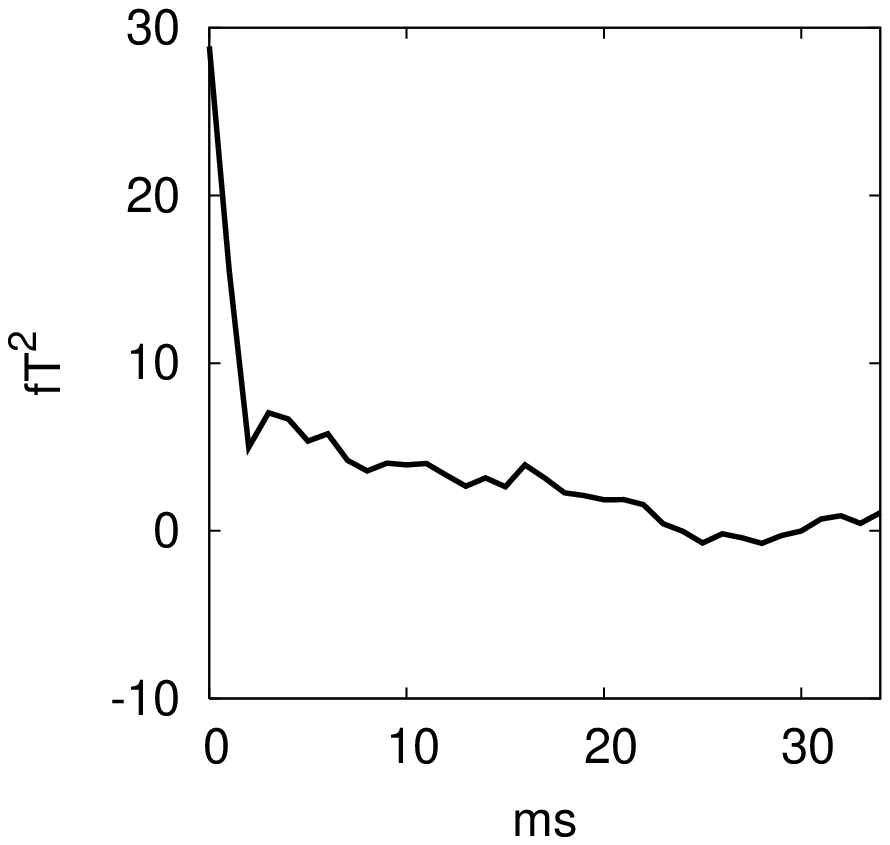} &
      \includegraphics[height = 1.25in, width = 1.25in]{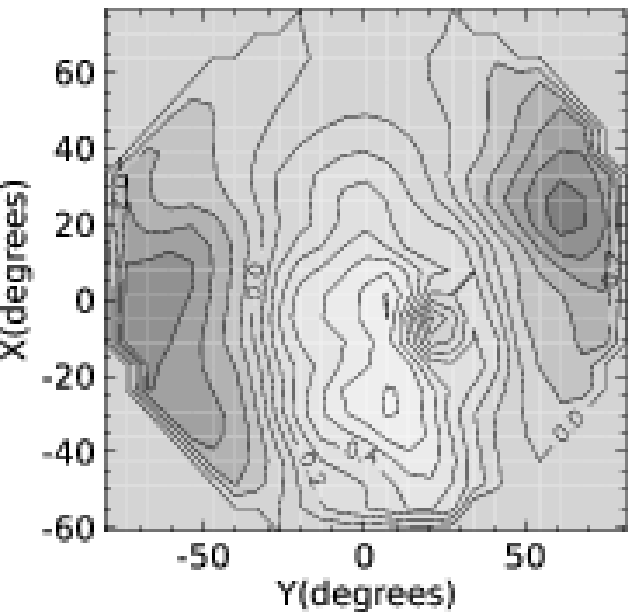}&
      \includegraphics[height = 1.25in]{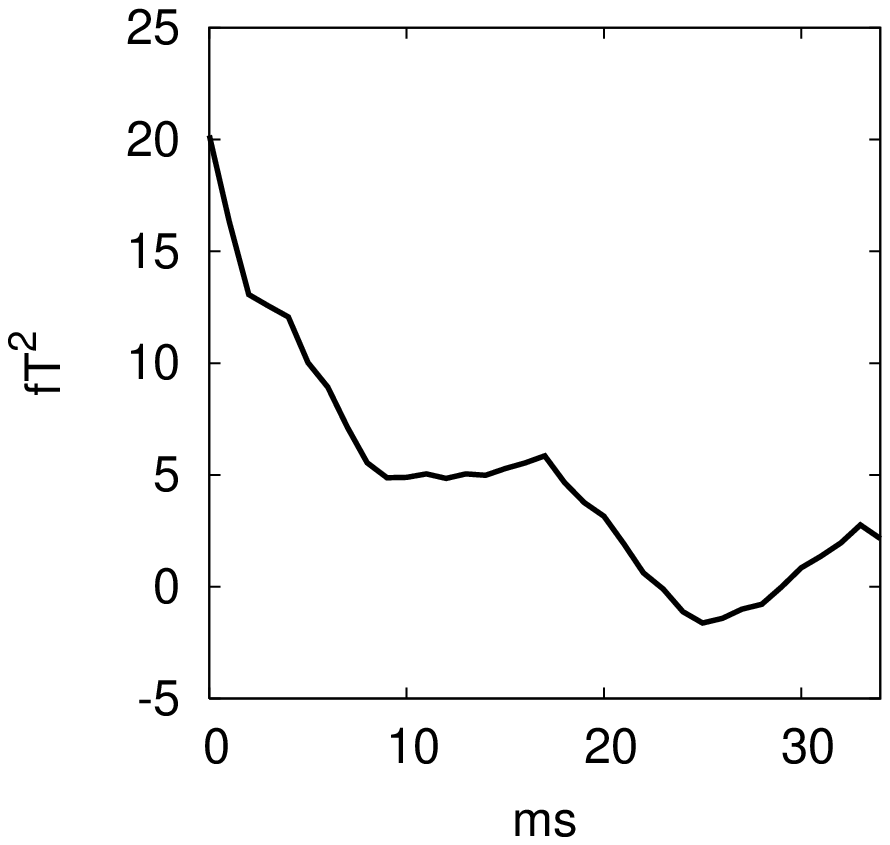}\\

      \raisebox{0.71in}{\tiny{\tt 3}}&
      \includegraphics[height = 1.25in, width = 1.25in]{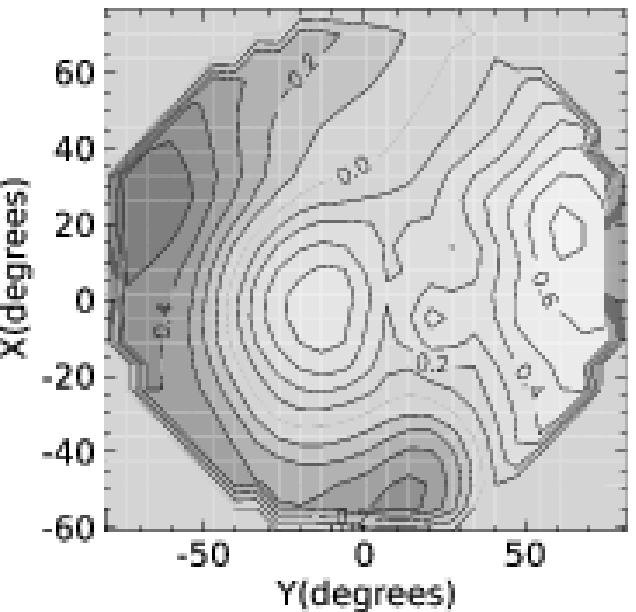}&
      \includegraphics[height = 1.25in]{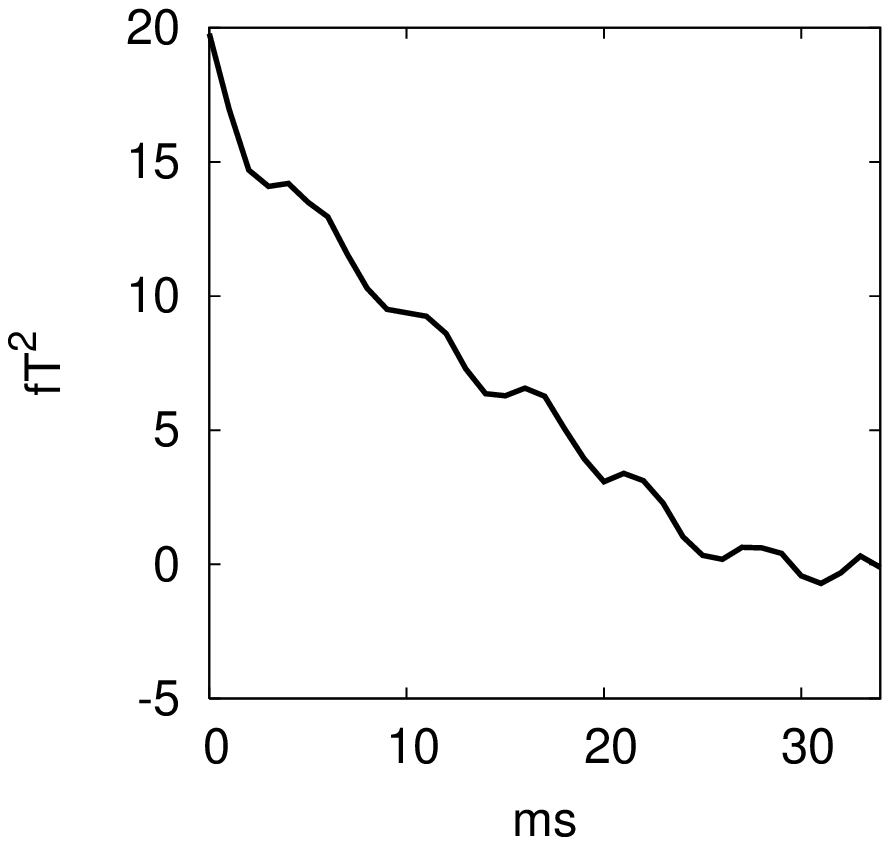}&
      \includegraphics[height = 1.25in, width = 1.25in]{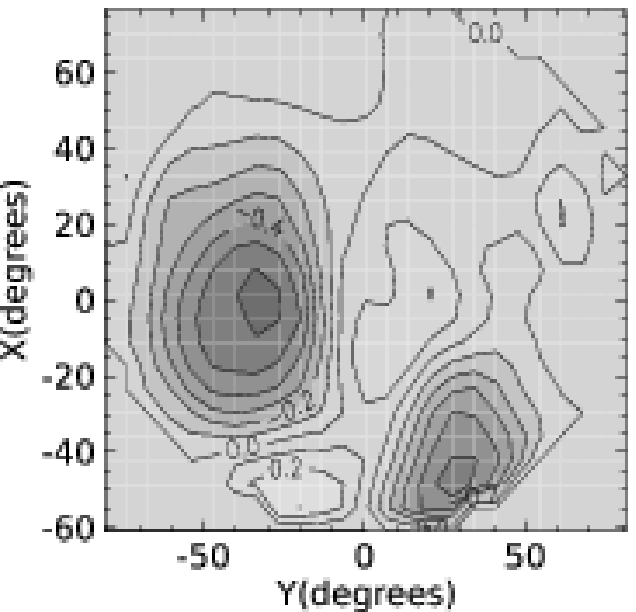} &
      \includegraphics[height = 1.25in]{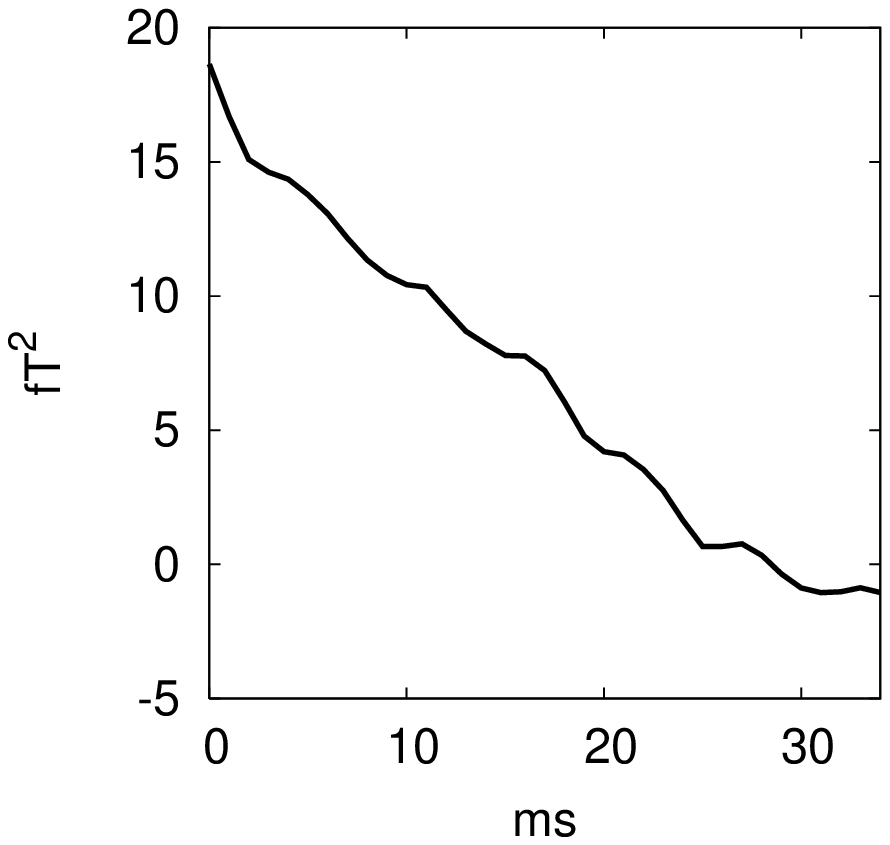} \\

      \raisebox{0.71in}{\tiny{\tt 4}}&
      \includegraphics[height = 1.25in, width = 1.25in]{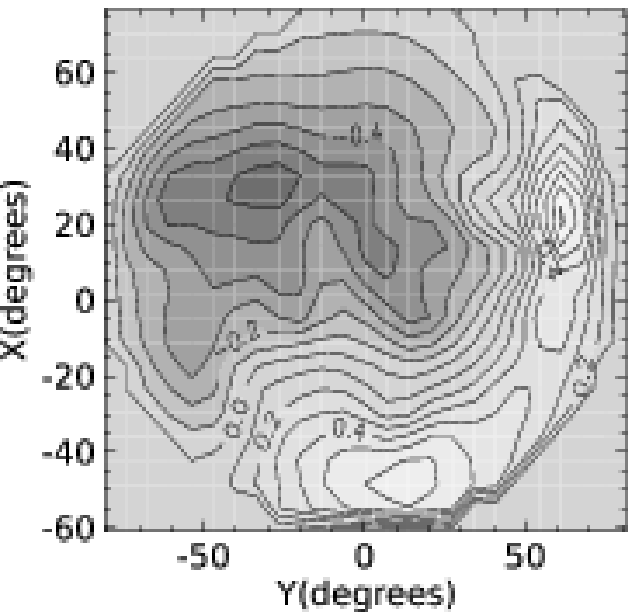}&
      \includegraphics[height = 1.25in]{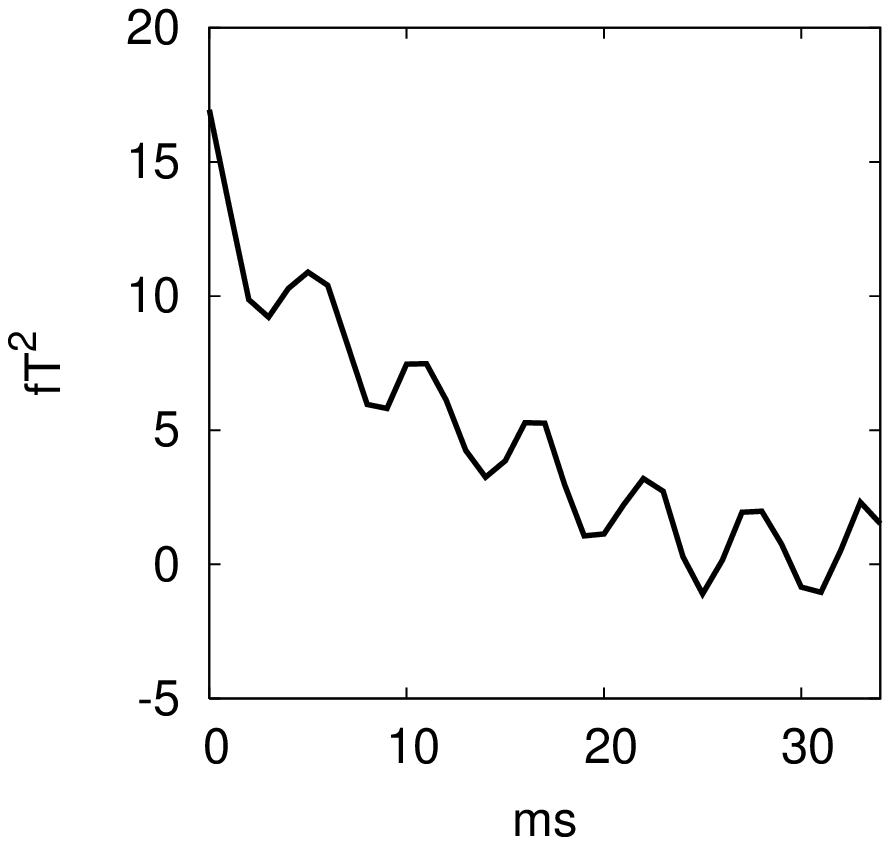} &
      \includegraphics[height = 1.25in, width = 1.25in]{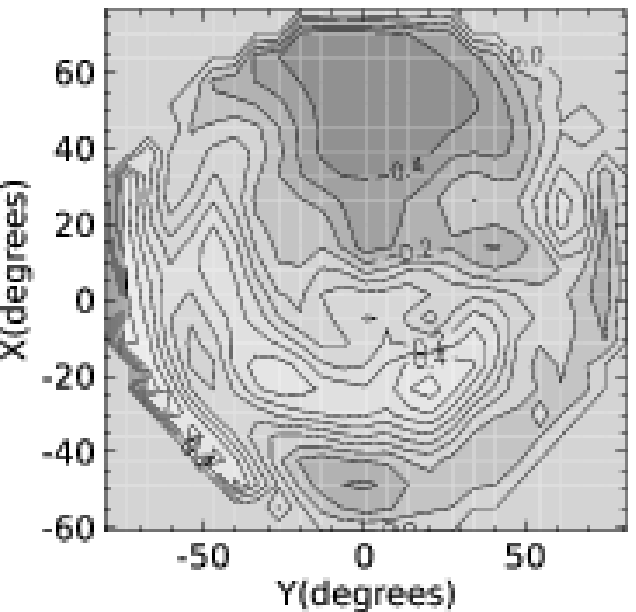} &
      \includegraphics[height = 1.25in]{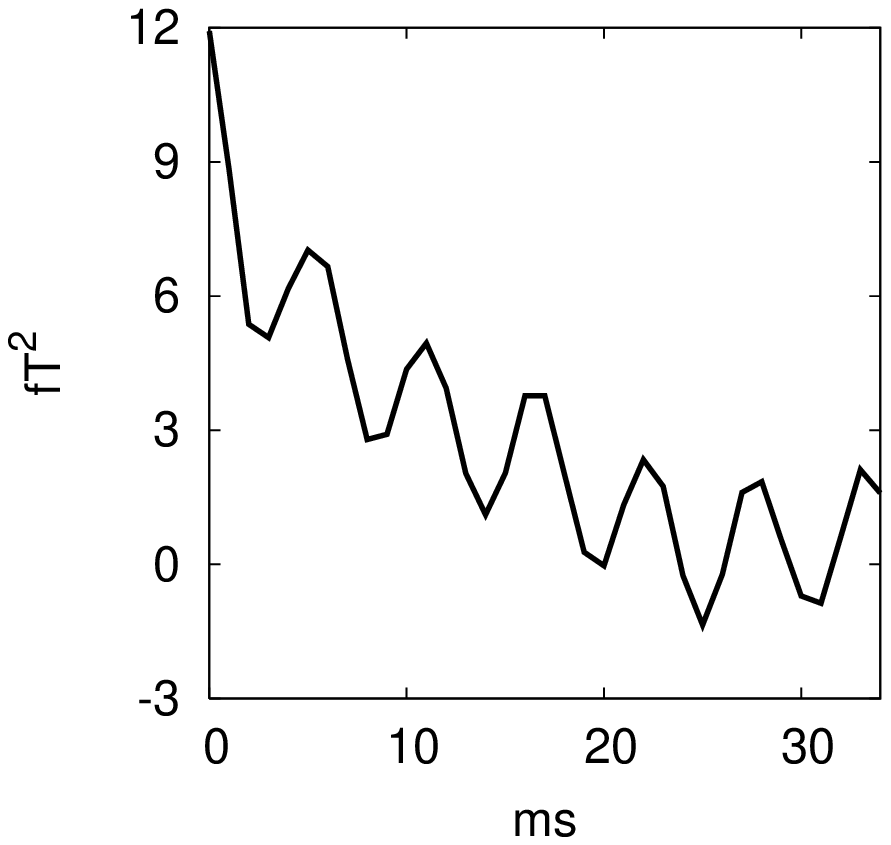}\\

      \raisebox{0.71in}{\tiny{\tt 5}}&
      \includegraphics[height = 1.25in, width = 1.25in]{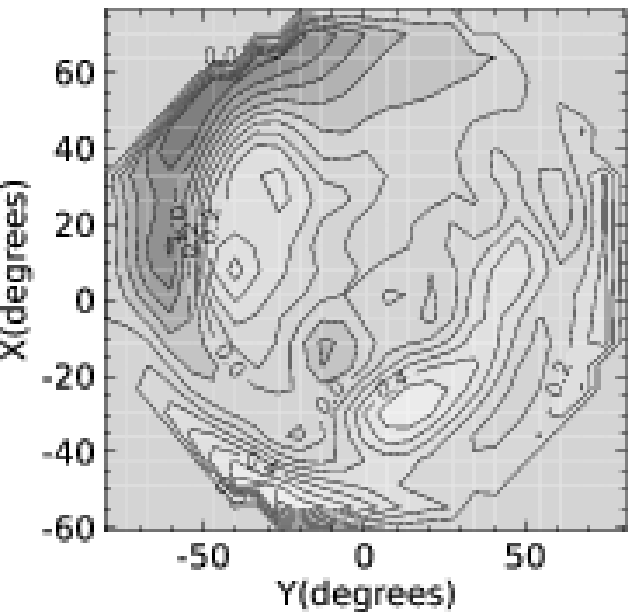}&
      \includegraphics[height = 1.25in]{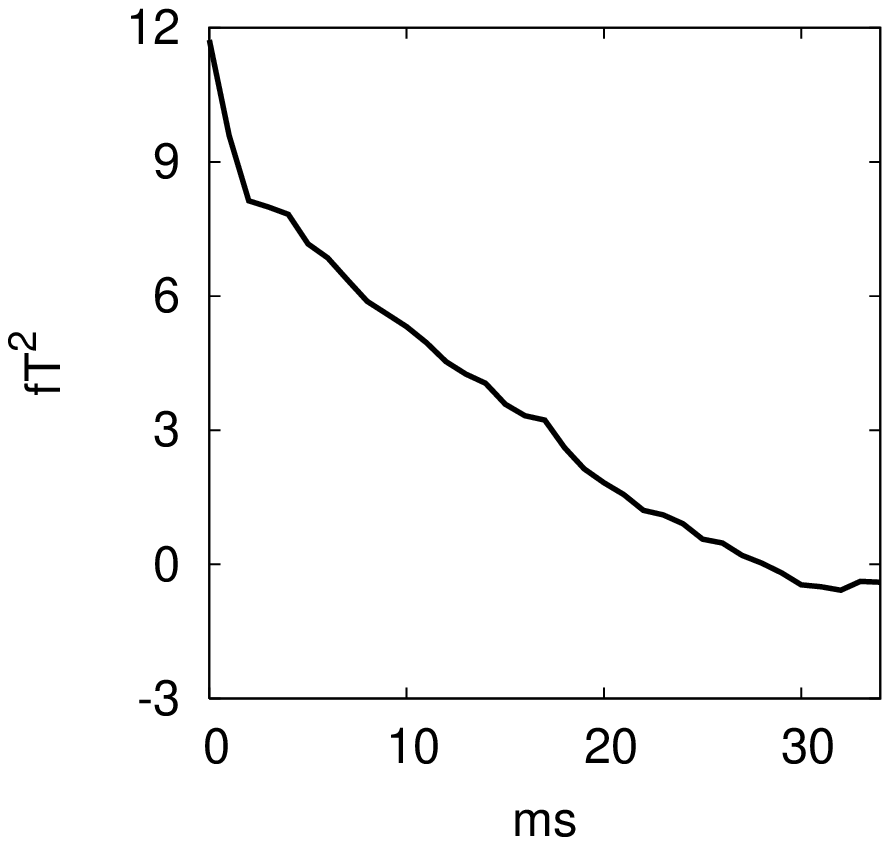}&
      \includegraphics[height = 1.25in, width = 1.25in]{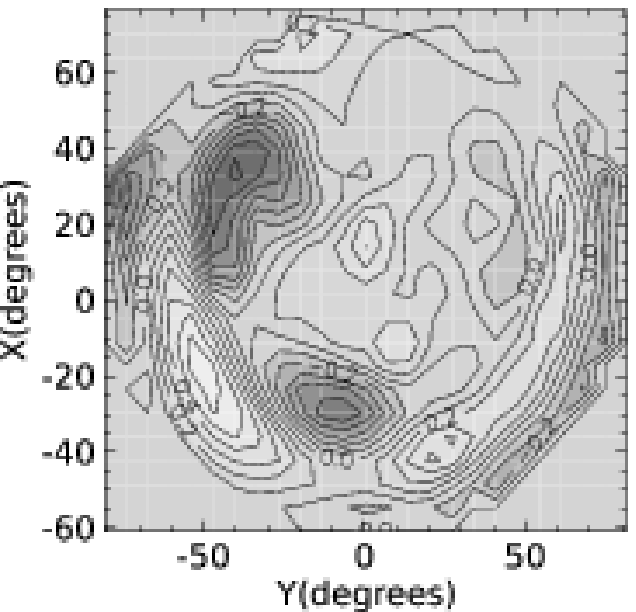}&
      \includegraphics[height = 1.25in]{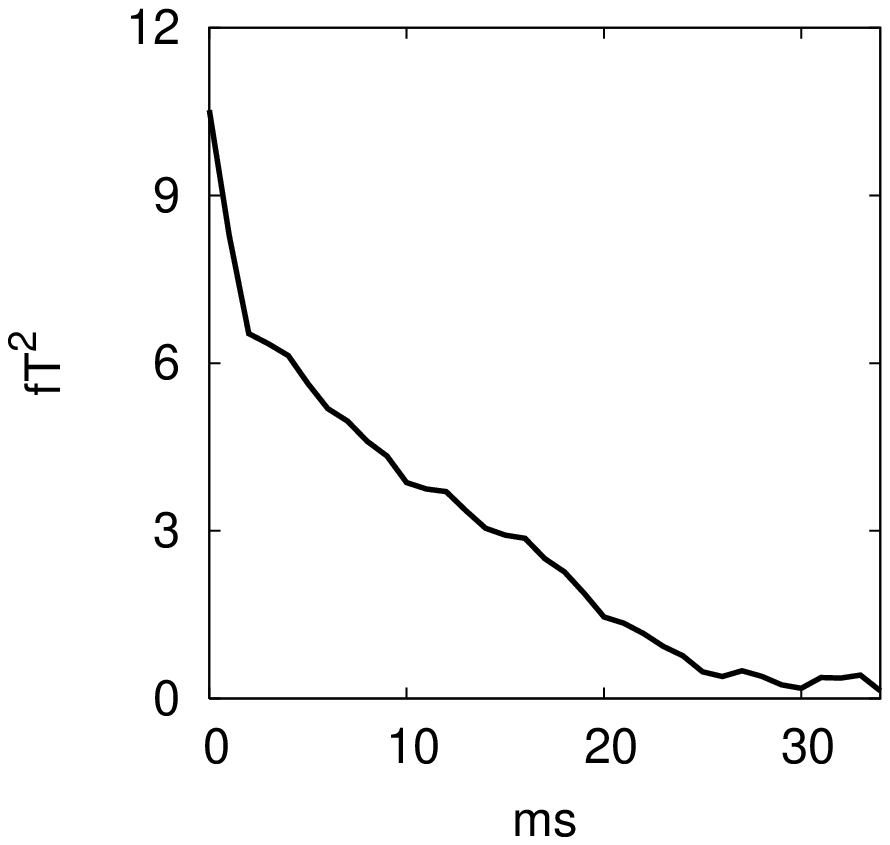}\\ 
      \cline{2-5}
      \raisebox{0.71in}{\tiny{\tt 6}}&
      \includegraphics[height = 1.25in, width = 1.25in]{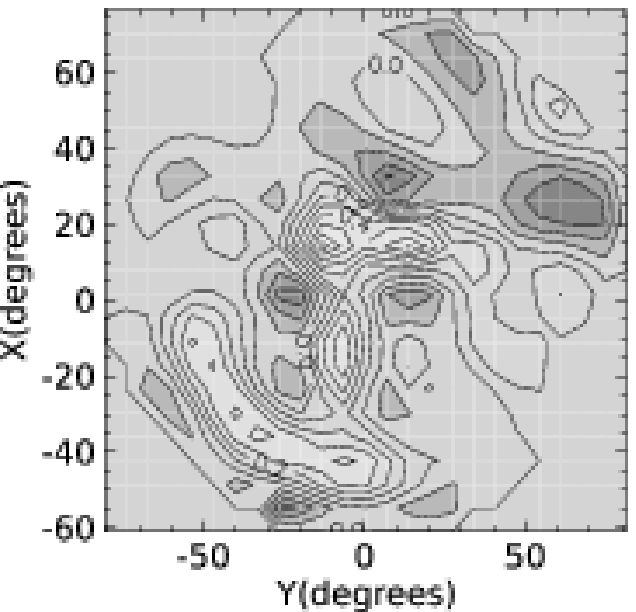}&
      \includegraphics[height = 1.25in]{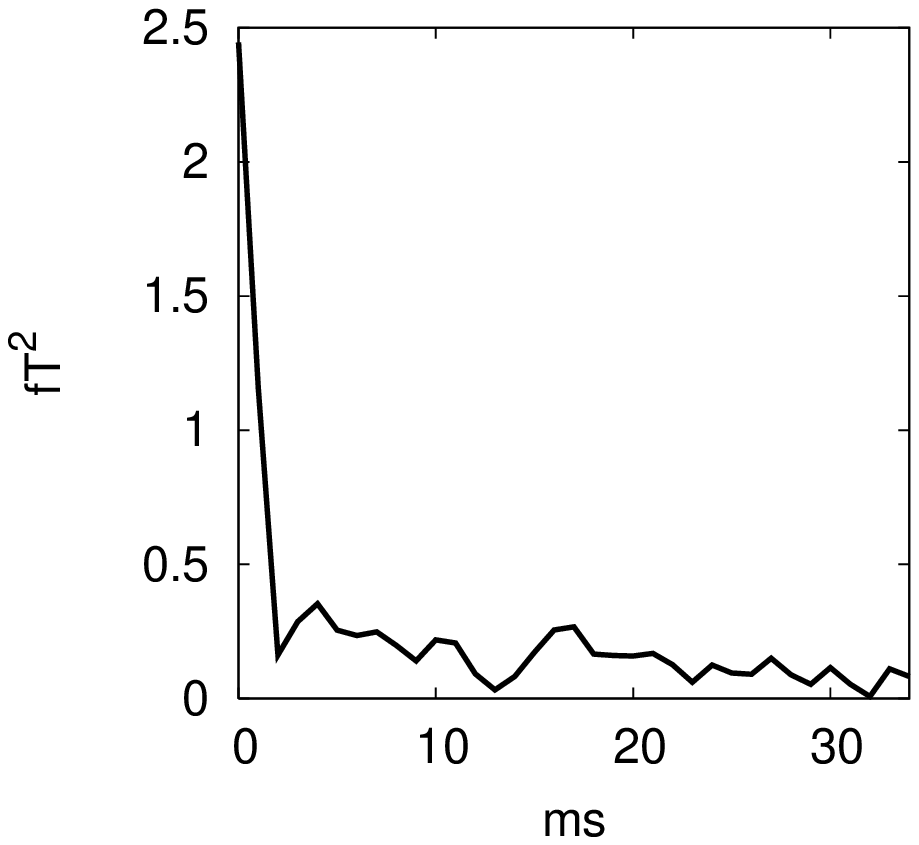}&
      \includegraphics[height = 1.25in, width = 1.25in]{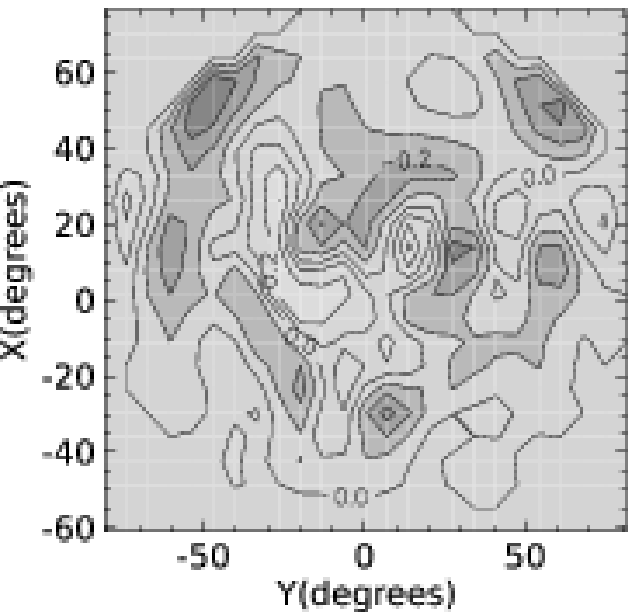}&
      \includegraphics[height = 1.25in]{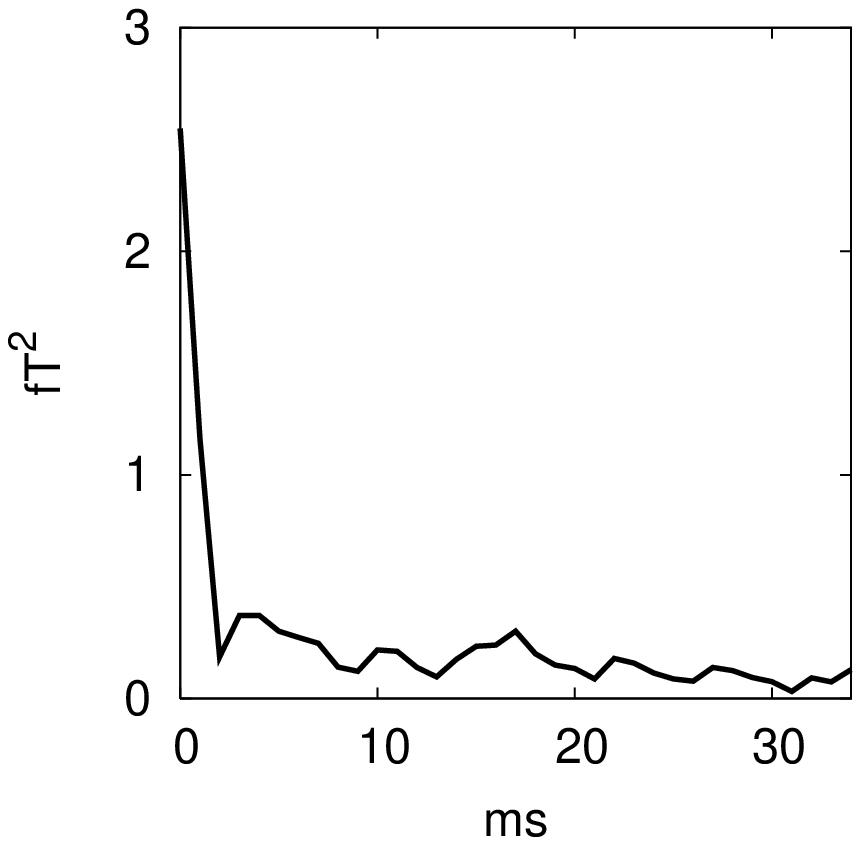}\\
      \cline{2-5}
      \multicolumn{2}{c}{{\tiny{\tt A}}}&{\tiny{\tt B}}&{\tiny{\tt C}}&
      \multicolumn{1}{c}{\tiny{\tt D}}
    \end{tabular}
  \end{center}
  \caption{Contour plots  of the first  10 the most  significant spatial
    components with  corresponding temporal covariance  functions (above
    the line) and of 2 the least significant spatial components.}
  \label{fig:helmets}
\end{figure}

This work,  like previous investigations,  underscores the value  of a
suitable  models  of  brain  and  other  noise  sources  for  accurate
localization  of neural  sources.   The most  sophisticated models  we
tested  gave the best  performance, with  improved estimation  of both
location and timecourse of the target neural source.

We also note the relatively small improvement in the multi-pair models
over  the single pair  models.  This  is especially  interesting given
that the  computational burden, both for estimating  the parameters of
the models  and in implementing  them in the inverse  calculations, is
relatively  small  for the  single  pair  model  and much  higher  for
multi-pair  models.  Thus  these findings  suggest that  a significant
gain in performance  may be had with a  relatively small computational
cost by  using a single  pair model.  However,  more work needs  to be
done to investigate the validity  and generality of this finding.  For
example, these results are from  a single subject, using single dipole
sources  and  over  a  limited  temporal  range.   Future  work  would
investigate whether  these results  hold over multiple  subjects, with
more complex  source configurations and  for a more  extended temporal
range. It is  possible that noise components will  be more variable or
more complex if these conditions change.

The assumption that each  noise generator contributing to the measured
background  does  not change  its  spatial  location  is supported  by
observations  of  stability  of  the  brain function  at  the  resting
state~\cite{RaichleME:defmbf}.   Spatial diversity  of  the background
signal   can  be   observed,  for   example,  as   nonuniform  spatial
distribution  of the  alpha  rhythm over  the  cortex.  Moreover,  the
background may not have many generators with distinct independent time
courses.  In the data  examined here  the temporal  covariances across
components were relatively similar  in the multi-pair models.  This is
a possible  explanation of why the multi-pair  models yield relatively
little improvement over the single pair models.

Figure  \ref{fig:helmets} shows  contour plots  of  spatial components
from the  SVD based multi-pair model and  their corresponding temporal
covariances.    There  are  clear   differences  among   the  temporal
covariances.  However, if one takes out the components that are either
sensor dominated with very  short temporal correlations (B2,B6 and D6)
and the large 60\%  powerline component (B1) the remaining components,
which  are presumably  dominated  by background  brain activity,  have
similar temporal correlation structure.  It will be interesting to see
if in  future work this  feature is observer across  multiple subjects
and over a more extended temporal range.

An observation  was made in the  course of this work  that may support
the supposition  that our data  does not exhibit great  variability in
the  hidden source  structure.  The  following heuristic  was  used to
estimate  one-pair Kronecker  model for  parallel comparison  with the
results presented in this paper. Suppose that each sensor has the same
temporal covariance.  Then it  is possible to estimate the single-pair
model using:
\begin{eqnarray}\label{eqn:temporal_simple}
  \hat{\mathbf{T}} &=& \frac{1}{M(LM-1)} \left( \mathbf{E}^{T}_{1},
    ... \mathbf{E}^{T}_{M} \right) 
  \left(
    \begin{array}{c}
      \mathbf{E}_{1} \\
      \vdots \\
      \mathbf{E}_{M}
    \end{array}
  \right) \nonumber \\
  \hat{\mathbf{S}} &=& \frac{1}{M(CM-1)} \left( \mathbf{E}_{1},
    ... \mathbf{E}_{M} \right) 
  \left( 
    \begin{array}{c}
      \mathbf{E}^{T}_{1} \\
      \vdots \\
      \mathbf{E}^{T}_{M}
    \end{array}
  \right),
  \label{eqn:mean}
\end{eqnarray}
where $\mathbf{E}_{m}$ is an $L\times  C$ single trial noise data with
$L$ number of sensors and $C$ number of time points; $M$ is the number
of trials. $\hat{\mathbf{T}}$ is  normalized to leave the variances in
$\hat{\mathbf{S}}$  only.   This is  not  an  optimized estimator,  as
compared   to   the    single-pair   Maximum   Likelihood   estimator;
nevertheless, we  found both of these  estimators performed similarly.
Looking at the two estimators, they could yield similar results if the
structure  of  the noise  is  not  very  complicated.  Furthermore,  a
compatible error might  be introduced to the MLE  based model when the
assumption about the Gaussian distribution of the un-averaged measured
background is violated.

Despite  almost  identical performance  of  the  multi-pair models  we
suggest that  in terms  of generality the  model based  on independent
basis should  be preferred. Its power  is not only in  the increase of
degrees of freedom  and consequently the ability to  describe a bigger
class of  possible spatial components  but also in  the discriminative
criterion.   The main  criterion used  in PCA  is minimization  of the
reconstruction  error~\cite{GERBRANDSJJ:THERBS}.   As  a  result  this
produces spatially  uncorrelated components.  This  may not be  a good
model of the MEG/EEG background, especially that due to the background
brain  activity.  It seems  more reasonable  to assume  that different
noise generators are independent.  In that case, the independent basis
model  is complex  enough to  describe  the underlying  process and  a
correct  choice  of  algorithm  (for  example an  ICA  algorithm)  can
estimate independent generators.

Considering the  observation that not all spatial  components may have
distinct temporal structure, a  further generalization is needed.  The
multi-pair model  can have an  additional parameter for the  number of
significant components. In this  case only significant components will
contribute to the final covariance with their respective distinct time
courses.  Other components would be suppressed. This generalization is
not obvious since special care  should be taken to retain inversion of
the model.  We leave this work for subsequent publications.

Multi-pair models can in principle be directly applied to the analysis
of EEG data.  Though the topic needs further study it is expected that
the different structure of EEG  would not affect the multi-pair models
as much as the possible added  complexity of the noise. We expect that
in the  case of heterogeneous  noise sources expected in  EEG signals,
multi-pair  models should show  even better  performance due  to their
increased  expressive  power and  ability  to incorporate  information
about many noise components.

We also  note that  a further reduction  of parameters  for multi-pair
models  can be achieved  by assuming  stationarity of  the background.
This  assumption  is  well  supported  by  studies,  see  for  example
\cite{BdM+03}.   Observation   of  time  courses   of  orthogonal  and
independent components for  the data used in this  study also supports
this  assumption.  Temporal   autocovariances  have  an  Toeplitz-like
structure  and   can  be   constructed  from  correlations   shown  in
Figure~\ref{fig:helmets}.   This  assumption  reduces  the  number  of
parameters to  be estimated to  $L(L+1)/2+LC$ for the SVD  based model
and  to  $L^{2}+LC$  for   the  ICA  based  model.  Another  important
improvement this assumption provides  is the reduction in running time
for  inverting   the  multi-pair  approximations   from  $O(LC^3)$  to
$O(LC^{2})$  \cite{Blahut:1985}. This will  be investigated  in future
work.

Different models that  would cover the range between  the diagonal and
the one-pair  model can  be developed. For  such models  an additional
study is  needed to investigate  how spatial models alone  or temporal
models alone  would compare in  performance as well.   These truncated
models can be useful in the case  when only a scarce amount of data is
available  for  estimation. But  when  the  single-pair  model can  be
estimated it  subsumes all lower  order models and should  not perform
worse than these.  As we  show in this paper multi-pair models subsume
the single-pair model and thus  should not perform worse than it. This
was demonstrated by the  increased localization accuracy.  In the case
when  sufficient  estimation  data  is available  and  the  additional
computational  load of  multi-pair  models does  not make  significant
difference, we  suggest using multi-pair models  to increase accuracy.
In the  worst case, multi-pair  models should not perform  poorer than
the one-pair model.

\section{Conclusions}
In  this  paper  we  have  introduced  two  multi-pair  spatiotemporal
covariance   models   as    a   generalization   over   the   previous
work~\cite{deMunck2002}  which   extends  the  expressive   power  and
provides better localization  results. The orthogonal basis multi-pair
model estimated using SVD is a more general but still a restricted one
due to  the orthogonality  constraints.  A more  general model  is the
independent basis  multi-pair model estimated using  an ICA algorithm.
Models  were  compared  on  the   basis  of  their  performance  in  a
localization  algorithm.  Performance  statistics  were gathered  from
inverse solutions  to a  large number of  single dipole  problems with
simulated  sources and  background from  real MEG  data.  In  terms of
localization  error,  the   orthogonal  basis  and  independent  basis
multi-pair models demonstrated  the best performance.  The independent
basis model is  a potentially better model due to  the increase in the
descriptive  power.  However, the  errors from  the single  pair model
examined were not much worse.   Future work will address whether these
findings  are a  general  feature of  MEG  background across  multiple
subjects,  with more  complex source  configurations and  over  a more
extended temporal range.

\section*{Acknowledgements}
This work was supported by NIH  grant 2 R01 EB000310-05 and the Mental
Illness  and Neuroscience  Discovery (MIND)  Institute. We  thank John
Mosher and Robert  Kraus for fruitful discussions and  help with noise
filtering. We thank Barak Pearlmutter for advice about ICA algorithms.
We thank  Elaine Best for  help in preprocessing empirical  data using
MEGAN                    (\url{http://www.lanl.gov/p/p21/megan.shtml}).
Figure~\ref{fig:dipoles}  showing  the   dipoles  on  the  cortex  was
generated using MRIVIEW~\cite{MRIVIEW-2002}.

\appendix
\section{Derivation of multi-pair models}
\label{sec:multi_pair_derivation}
For the following derivation we  need to calculate determinants of the
multi-pair models.  This is done in the Appendix~\ref{sec:determinant}
for the  case of orthogonal basis  model~(\ref{eq:svd_det}).  Also the
inverse of the  orthogonal basis multi-pair model~(\ref{eq:ortho_inv})
is utilized  in the following.  Using these result the  log likelihood
for  the Gaussian  pdf for  the orthogonal  basis multi-pair  model is
expressed like:
\begin{equation}
  {\mathcal L} = const-\frac{M}{2}\sum_{l=1}^{L} 
  \ln(|\bm{\mathcal T}^{l}|) - 
  \frac{1}{2}tr(\sum_{m=1}^{M} \sum_{l=1}^{L} 
  \mathbf{E}_{m}^{T}  \bm{{\mathcal S}}^l \mathbf{E}_{m} 
  (\bm{\mathcal T}^{l})^{-1})
\end{equation}

Differentiating with respect to $\bm{\mathcal T}^{l}$ results in:
\begin{eqnarray} 
  d{\mathcal L} &=& -\frac{M}{2}tr((\bm{\mathcal T}^{l})^{-1} 
  d\bm{\mathcal T}^{l}) + 
  \frac{1}{2}tr(\sum_{m=1}^{M}\mathbf{E}_{m}^{T} \bm{{\mathcal S}}^l
  \mathbf{E}_{m} 
  (\bm{\mathcal T}^{l})^{-1} d\bm{\mathcal T}^{l} 
  (\bm{\mathcal T}^{l})^{-1}) 
  \nonumber \\
  &=& -\frac{M}{2}tr((\bm{\mathcal T}^{l})^{-1}d\bm{\mathcal T}^{l}) + 
  \frac{1}{2}tr(\sum_{m=1}^{M} (\bm{\mathcal T}^{l})^{-1} 
  \mathbf{E}_{m}^{T} 
  \bm{{\mathcal S}}^l \mathbf{E}_{m}
  (\bm{\mathcal T}^{l})^{-1} d\bm{\mathcal T}^{l}) \nonumber \\
  &=& -\frac{M}{2}tr([ (\bm{\mathcal T}^{l})^{-1} - 
  \frac{1}{M}\sum_{m=1}^{M}(\bm{\mathcal T}^{l})^{-1} \mathbf{E}_{m}^{T} 
  \bm{{\mathcal S}}^l \mathbf{E}_{m}
  (\bm{\mathcal T}^{l})^{-1}] d\bm{\mathcal T}^{l})
\end{eqnarray}
And the final result is:
\begin{equation}
  \bm{\mathcal T}^{l} = \frac{1}{M}\sum_{m=1}^{M}\mathbf{E}_{m}^{T} 
  \bm{{\mathcal S}}^l \mathbf{E}_{m} 
\end{equation}

Next  we  estimate  temporal  covariance  for  the  independent  basis
multi-pair model.   The determinant of the independent  basis model is
calculated              in              Appendix~\ref{sec:determinant}
equation~(\ref{eq:ica_det}) and the inverse  of this model is taken as
in~(\ref{eq:ica_inv}).  The log likelihood in this case:
\begin{eqnarray}
  {\mathcal L} &=& const-\frac{M}{2}(2C\ln(|\mathbf{W}^{-1}|) + 
  \sum_{l=1}^{L}\ln(|\bm{\mathcal T}^{l}|)) \nonumber\\ &&- 
  \frac{1}{2}tr(\sum_{m=1}^{M} \sum_{l=1}^{L} 
  \mathbf{E}_{m}^{T} \mathbf{WW}^{T}\bm{\mathcal R}^{l} \mathbf{WW}^{T}
  \mathbf{E}_{m} (\bm{\mathcal T}^{l})^{-1})
\end{eqnarray}

Following the  same path  for derivation as  for the  orthogonal basis
model case we end up with:
\begin{equation}
  \bm{\mathcal T}^{l} = \frac{1}{M}\sum_{m=1}^{M}\mathbf{E}_{m}^{T} 
  \mathbf{WW}^{T}\bm{\mathcal R}^{l}\mathbf{WW}^{T} \mathbf{E}_{m}
  \label{eq:ica_mle}
\end{equation}

\section{Inversion of the multi-pair models}
\label{sec:multi_pair_inversion}
We need the following identity for derivations of this section:
\begin{equation}
  (\mathbf{A}_1 \otimes \mathbf{B}_1 )
  (\mathbf{A}_2 \otimes \mathbf{B}_2) = (\mathbf{A}_1 \mathbf{A}_2
  \otimes \mathbf{B}_1
  \mathbf{B}_2)\label{eqn:kroneker_identity}
\end{equation} 

First, lets prove that the inverse for the orthogonal basis multi-pair
model  is  as  in~(\ref{eq:ortho_inv})   .  By  the  orthogonality  of
$\bm{{\mathcal S}}^l$  ($\bm{{\mathcal S}}^l \bm{{\mathcal  S}}^{l'} =
0$ if $ l \ne {l'} $),
\begin{eqnarray}\nonumber
  \widetilde{\mathbf{COV}}\ \widetilde{\mathbf{COV}}^{-1} &=& \left (\sum_{l=1}^L 
    \lambda_l^2 \bm{\mathcal T}^l \otimes \bm{{\mathcal S}}^l
  \right ) \left (\sum_{l=1}^L \lambda_l^{-2} (\bm{\mathcal T}^l)^{-1}
    \otimes \bm{{\mathcal S}}^l \right )\\
  &=& \sum_{l=1}^L \mathbf{I} \otimes (\bm{{\mathcal S}}^l)^2 
\end{eqnarray}
By two properties of $\bm{{\mathcal S}}^l$: (1) $(\bm{{\mathcal S}}^l)^2 =
\bm{{\mathcal S}}^l$; (2) $\sum_l \bm{{\mathcal S}}^l = \mathbf{I}$ due to
orthogonality of $\mathbf{V}$, we obtain the desired one:
\begin{equation}
  \widetilde{\mathbf{COV}}\ \widetilde{\mathbf{COV}}^{-1} =  
  \sum_{l=1}^L \mathbf{I} \otimes (\bm{{\mathcal S}}^l)^2 = \mathbf{I}
  \otimes \sum_{l=1}^L \bm{{\mathcal S}}^l = \mathbf{I} \otimes \mathbf{I} =
  \mathbf{I}.
\end{equation}

Properties of $\bm{{\mathcal S}}^l$ are proved as follows: 
\begin{itemize}
\item property (1) 
  \begin{eqnarray}\nonumber
    (\bm{{\mathcal S}}^l)^2 &=& (v_l v_l^T) (v_l v_l^T) = v_l (v_l^T v_l) v_l^T
    \\
    \nonumber
    &=& v_l v_l^T \quad ( v_l^T v_l = 1 \mbox{ by orthogonality of $\mathbf{V}$})
    \\
    &=& \bm{{\mathcal S}}^l
  \end{eqnarray}
\item property (2) :
  $
  \mathbf{I} = \mathbf{V} \ \mathbf{V}^T = \sum_l v_l v_l^T 
  $
\end{itemize}

Now we show that the  inverse of the independent components multi-pair
model       is      as      in~(\ref{eq:ica_inv})       Using      the
identity~(\ref{eqn:kroneker_identity})
\begin{eqnarray}\nonumber
  & & \left (\sum_{l=1}^L  \bm{\mathcal T}^l \otimes \bm{\mathcal R}^l
  \right ) \left (\sum_{l=1}^L (\bm{\mathcal T}^l)^{-1}
    \otimes [\mathbf{WW}^T \bm{\mathcal R}^l \mathbf{WW}^T] \right )\\
  & & \quad = \sum_{l, l'=1}^L \bm{\mathcal T}^l (\bm{\mathcal
    T}^{l'})^{-1} \otimes [\bm{\mathcal R}^l \mathbf{W W}^T \bm{\mathcal
    R}^{l'} \mathbf{W W}^T].
  \label{eqn:ICA_inverse_initial}
\end{eqnarray}
According to the definition of $\bm{\mathcal R}_l$:
\begin{equation}
  \label{eqn:spatial_cov_from_IC}
  \bm{\mathcal R}_l = w_l w_l^T = (\mathbf{W}^{-1})^T e_l e_l^T
  \mathbf{W}^{-1}.
\end{equation}
Here $\{e_l| l = 1, \cdots, L\}$ are orthonormal canonical bases
vectors. By the orthogonality of $e_l$ and
substituting~(\ref{eqn:spatial_cov_from_IC}) into
(\ref{eqn:ICA_inverse_initial}), we obtain
\begin{eqnarray}   \nonumber    \bm{\mathcal   R}^l   \mathbf{W   W}^T
  \bm{\mathcal  R}^{l'} \mathbf{W W}^T &=& (\mathbf{W}^{-1})^T e_l
  e_l^T \mathbf{W}^{-1} \mathbf{W W}^T
  (\mathbf{W}^{-1})^T e_{l'} e_{l'}^T \mathbf{W}^{-1} \mathbf{W W}^T \\
  \nonumber 
  &=& (\mathbf{W}^{-1})^T e_l e_l^T e_{l'} e_{l'}^T \mathbf{W}^T \\
  &=&
  \delta(l,l') (\mathbf{W}^{-1})^T e_l e_l^T \mathbf{W}^T
\end{eqnarray}
Finally, we obtain the desired result:
\begin{eqnarray}\nonumber
  & & \left (\sum_{l=1}^L  \bm{\mathcal T}^l \otimes \bm{\mathcal R}^l
  \right ) \left (\sum_{l=1}^L (\bm{\mathcal T}^l)^{-1}
    \otimes [\mathbf{W W}^T \bm{\mathcal R}^l \mathbf{W W}^T] \right )\\
  & & \quad = \mathbf{I}  \otimes (\mathbf{W}^{-1})^T [\sum_l e_l e_l^T]
  \mathbf{W}^T = \mathbf{I} \otimes (\mathbf{W}^{-1})^T \mathbf{I}
  \mathbf{W}^T = \mathbf{I}.
\end{eqnarray}

\section{Calculation of the determinant}
\label{sec:determinant}

Before performing actual calculations of the determinant we transfrom
the original series form to a shape more manageable in our further
reasoning. The only assumption needed for the transformation is that
matrices $S^{l}$ are rank one outer products $v_{l}v_{l}^{T}$, where
$v_{l}$ comes from an independent basis set. 

We first rewrite $\displaystyle\sum T^{l} \otimes S^{l}$ in the matrix
form:
\begin{equation}
  \displaystyle\sum T^{l} \otimes S^{l} = \left[
    \begin{array}{cccc}
      \sum_{l=1}^{L} T^{l}_{1,1}  S^{l} & 
      \sum_{l=1}^{L} T^{l}_{1,2}  S^{l} & \cdots & 
      \sum_{l=1}^{L} T^{l}_{1,C}  S^{l} \\ 
      \sum_{l=1}^{L} T^{l}_{2,1}  S^{l} & 
      \sum_{l=1}^{L} T^{l}_{2,2}  S^{l} & \cdots & 
      \sum_{l=1}^{L} T^{l}_{2,C}  S^{l} \\ 
      \vdots & \vdots & \ddots & \vdots \\
      \sum_{l=1}^{L} T^{l}_{C,1}  S^{l} & 
      \sum_{l=1}^{L} T^{l}_{C,2}  S^{l} & \cdots & 
      \sum_{l=1}^{L} T^{l}_{C,C}  S^{l} \\ 
    \end{array}
  \right]
\end{equation}
Considering each element separately we can see that:
\begin{equation}
  \sum_{l=1}^{L} T^{l}_{i,j} S^{l} = \sum_{l=1}^{L} T^{l}_{i,j} v_{l}v_{l}^T
  = S \Lambda_{i,j} S^{T},
  \label{eq:diagonalization}
\end{equation}
where $S$ is an $L\times L$ matrix with columns being $v_{l}$ vectors
from an orthonormal basis; $\Lambda_{i,j}$ is a diagonal matrix with
the elements along the diagonal being $\{T^{1}_{i,j}, T^{2}_{i,j},
\ldots, T^{L}_{i,j}\}$. Observing~(\ref{eq:diagonalization}) one more
rewrite is possible:
\begin{equation}
  \displaystyle\sum T^{l} \otimes S^{l} =
  \left[     
    \begin{array}{cccc}
      S & 0 & \cdots & 0 \\ 
      0 & S & \cdots & 0 \\ 
      \vdots & \vdots & \ddots & \vdots \\
      0 & 0 & \cdots & S \\ 
    \end{array}
  \right]
  \left[     
    \begin{array}{cccc}
      \Lambda_{1,1} & \Lambda_{1,2} & \cdots & \Lambda_{1,C} \\ 
      \Lambda_{2,1} & \Lambda_{2,2} & \cdots & \Lambda_{2,C} \\ 
      \vdots & \vdots & \ddots & \vdots \\
      \Lambda_{C,1} & \Lambda_{C,2} & \cdots & \Lambda_{C,C} \\ 
    \end{array}
  \right]
  \left[     
    \begin{array}{cccc}
      S^{T} & 0 & \cdots & 0 \\ 
      0 & S^{T} & \cdots & 0 \\ 
      \vdots & \vdots & \ddots & \vdots \\
      0 & 0 & \cdots & S^{T} \\ 
    \end{array}
  \right]
\end{equation}

The next step is actually calculating the determinant
$\det\left(\displaystyle\sum T^{l} \otimes S^{l}\right)$:
\begin{equation}
  \det\left(\left[     
      \begin{array}{cccc}
        S & 0 & \cdots & 0 \\ 
        0 & S & \cdots & 0 \\ 
        \vdots & \vdots & \ddots & \vdots \\
        0 & 0 & \cdots & S \\ 
      \end{array}
    \right]\right)
  \det\left(\left[     
      \begin{array}{cccc}
        \Lambda_{1,1} & \Lambda_{1,2} & \cdots & \Lambda_{1,C} \\ 
        \Lambda_{2,1} & \Lambda_{2,2} & \cdots & \Lambda_{2,C} \\ 
        \vdots & \vdots & \ddots & \vdots \\
        \Lambda_{C,1} & \Lambda_{C,2} & \cdots & \Lambda_{C,C} \\ 
      \end{array}
    \right]\right)
  \det\left(\left[     
      \begin{array}{cccc}
        S^{T} & 0 & \cdots & 0 \\ 
        0 & S^{T} & \cdots & 0 \\ 
        \vdots & \vdots & \ddots & \vdots \\
        0 & 0 & \cdots & S^{T} \\ 
      \end{array}
    \right]\right)
  \label{eq:3det}
\end{equation}

Let's  leave  diagonal matrices  alone  since  their determinants  are
trivial and calculate the middle determinant. Let's introduce $E^{l} =
e_{l}e_{l}^{T}$, where  $e_{l}$ is a  basis vector from  the canonical
basis  set. Now the  matrix of  ${\Lambda_{i,j}}$ can  be consequently
rewritten as:
\begin{equation}
  \displaystyle \sum_{l=1}^{L} T^{l} \otimes E^{l} = \left[
    \begin{array}{cccc}
      \sum_{l=1}^{L} T^{l}_{1,1}  E^{l} & 
      \sum_{l=1}^{L} T^{l}_{1,2}  E^{l} & \cdots & 
      \sum_{l=1}^{L} T^{l}_{1,C}  E^{l} \\ 
      \sum_{l=1}^{L} T^{l}_{2,1}  E^{l} & 
      \sum_{l=1}^{L} T^{l}_{2,2}  E^{l} & \cdots & 
      \sum_{l=1}^{L} T^{l}_{2,C}  E^{l} \\ 
      \vdots & \vdots & \ddots & \vdots \\
      \sum_{l=1}^{L} T^{l}_{C,1}  E^{l} & 
      \sum_{l=1}^{L} T^{l}_{C,2}  E^{l} & \cdots & 
      \sum_{l=1}^{L} T^{l}_{C,C}  E^{l} \\ 
    \end{array}
  \right]
\end{equation}

By definition the determinant of a $C\times C$ matrix $A$ is:
\begin{equation}
  \det(A) = \sum_{\rho} \sigma(\rho) 
  a_{1,\rho_{1}}a_{1,\rho_{2}}\ldots a_{1,\rho_{C}},
\end{equation}
where $\rho$ is a permutation, $\sigma(\rho)$ is the sign of a
permutation and $a_{i,j}$ is an element of $A$.

Because of the orthogonality of $E^{l}$ matrices to each other when
their indices are not the same and using the property
$(E^{l})^{n}=E^{l}$:
\begin{eqnarray}
  \det\left(\displaystyle \sum_{l=1}^{L} T^{l} \otimes E^{l}\right) &=& 
  \det\left(\displaystyle \sum_{\rho}\sigma(\rho) \displaystyle \sum_{l}
    T^{l}_{1,\rho_{1}}T^{l}_{1,\rho_{2}}\ldots T^{l}_{1,\rho_{C}} E^{l} 
  \right)\\
  \det\left(\displaystyle \sum_{l=1}^{L} T^{l} \otimes E^{l}\right) &=& 
  \det\left(\displaystyle \sum_{l} \left( \displaystyle \sum_{\rho}\sigma(\rho)
      T^{l}_{1,\rho_{1}}T^{l}_{1,\rho_{2}}\ldots T^{l}_{1,\rho_{C}}\right)
    E^{l} \right)\\
  \det\left(\displaystyle \sum_{l=1}^{L} T^{l} \otimes E^{l}\right) &=& 
  \det\left(\displaystyle \sum_{l} \det\left(T^{l}\right)
    E^{l} \right)\\
  \det\left(\displaystyle \sum_{l=1}^{L} T^{l} \otimes E^{l}\right) &=& 
  \det\left(I \left[ 
      \begin{array}{ccc}
        \ddots &  & 0 \\
        & \det\left(T^{l}\right) & \\
        0 &  & \ddots \\
      \end{array} \right]
    I \right)\\
  \det\left(\displaystyle \sum_{l=1}^{L} T^{l} \otimes E^{l}\right) &=& 
  \displaystyle \prod_{l}^{L} \det\left(T^{l}\right)
  \label{eq:proddet}
\end{eqnarray}

With this result~(\ref{eq:proddet}) the determinant
from~(\ref{eq:3det}) can be rewritten:
\begin{equation}
  \det\left(\displaystyle\sum T^{l} \otimes S^{l}\right) = 
  \det\left(S\right)^{2C}  \displaystyle \prod_{l}^{L} \det\left(T^{l}\right)
  \label{eq:general_det}
\end{equation}

We first calculate a special case when $\bm{\mathcal S}^{l} =
v_{l}v_{l}^T$ and all column vectors $v_{l}$ belong to an orthonormal
basis, that is $S$ from~(\ref{eq:general_det}) in an orthogonal
matrix. In this case $\det(S) = 1$ and the result is:
\begin{equation}
  \det\left(\displaystyle\sum T^{l} \otimes S^{l}\right) = 
  \displaystyle \prod_{l}^{L} \det\left(T^{l}\right)
  \label{eq:svd_det}
\end{equation}

In the case when $S$ is not orthogonal like in the ICA case, we get
slightly different result. Redefine $S^{l} = R^{l}$, where $R^{l} =
w_{l}w_{l}^{T}$ and $w_{l}$ is a row vector of the mixing matrix
$W^{-1}$. Then~(\ref{eq:general_det}) becomes:
\begin{equation}
  \det\left(\displaystyle\sum T^{l} \otimes R^{l}\right) = 
  \det\left(W^{-1}\right)^{2C}  
  \displaystyle \prod_{l}^{L} \det\left(T^{l}\right)
  \label{eq:ica_det}
\end{equation}

\bibliographystyle{elsart-num}
\bibliography{Covariance-Neuroimage}

\end{document}